\documentclass[11pt]{article}

\usepackage[pdftex]{graphicx}
\graphicspath{{figs/}{figsnew/}} 

\usepackage{hyperref}

\usepackage[T1]{fontenc}
\usepackage{dfadobe}  
\usepackage{macrosetc}
\usepackage{cite}

\usepackage{xcolor}

\usepackage{amssymb}
\usepackage{amsmath}
\usepackage{amsfonts}

\newcommand{\mapper}{\mm{\mathcal M}}
\newcommand{\Vcal}        {\mm{\mathcal V}}

\definecolor{darkgrn}{rgb}{0, 0.75, 0}

\newcommand{\tool} {\mbox{{\textsf{Pheno-Mapper}}}}

\newcommand{\etal} {{\textit{et al.}}}

\usepackage{mathptmx}
\DeclareMathAlphabet{\mathcal}{OMS}{cmsy}{m}{n}
\usepackage{mathtools}
\DeclareMathDelimiter{(}{\mathopen} {operators}{"28}{largesymbols}{"00}
\DeclareMathDelimiter{)}{\mathclose}{operators}{"29}{largesymbols}{"01}

\usepackage{tabularx}

\usepackage{xcolor,colortbl}
\definecolor{green}{rgb}{0.1,0.1,0.1}

\definecolor{amethyst}{rgb}{0.6, 0.4, 0.8}
\definecolor{aliceblue}{rgb}{0.94, 0.97, 1.0}
\definecolor{apricot}{rgb}{0.98, 0.81, 0.69}
\definecolor{aquamarine}{rgb}{0.5, 1.0, 0.83}	
\definecolor{ashgrey}{rgb}{0.7, 0.75, 0.71}
\definecolor{asparagus}{rgb}{0.53, 0.66, 0.42}
\definecolor{babyblue}{rgb}{0.54, 0.81, 0.94}
\definecolor{babypink}{rgb}{0.96, 0.76, 0.76}
\definecolor{burlywood}{rgb}{0.87, 0.72, 0.53}
\definecolor{brightlavender}{rgb}{0.75, 0.58, 0.89}

\providecommand{\keywords}[1]
{
  \small	
  \textbf{\textit{Keywords---}} #1
}

\begin{document}

\title{Pheno-Mapper: An Interactive Toolbox for the Visual Exploration of Phenomics Data}

\author{
     Youjia Zhou            \thanks{University of Utah, zhou325@sci.utah.edu}
\and Methun Kamruzzaman      \thanks{University of Virginia, hkz8wk@virginia.edu}
\and Patrick Schnable   \thanks{Iowa State University,  schnable@iastate.edu} 
\and Bala Krishnamoorthy          \thanks{Washington State University,  kbala@wsu.edu}
\and Ananth Kalyanaraman \thanks{Washington State University, ananth@wsu.edu}
\and Bei Wang               \thanks{University of Utah, beiwang@sci.utah.edu.}
}

\date{}   


\maketitle

\begin{abstract}
High-throughput technologies to collect field data have made observations possible at scale in several branches of life sciences. 
The data collected can range from the molecular level (genotypes) to physiological  (phenotypic traits) and environmental observations (e.g., weather, soil conditions). 
These vast swathes of data, collectively referred to as \emph{phenomics} data, represent a treasure trove of key scientific knowledge on the dynamics of the underlying biological system. 
However, extracting information and insights from these complex datasets remains a significant challenge owing to their multidimensionality and lack of prior knowledge about their complex structure.  
In this paper, we present {\tool}, an interactive toolbox for the exploratory analysis and visualization of large-scale phenomics data. 
Our approach uses the \emph{mapper} framework to perform a topological analysis of the data, and subsequently render visual representations with built-in data analysis and machine learning capabilities. 
We demonstrate the utility of this new tool on real-world plant (e.g., maize)  phenomics datasets. 
In comparison to existing approaches, the main advantage of {\tool} is that it provides rich, interactive capabilities in the exploratory analysis of phenomics data, and it integrates visual analytics with data analysis and machine learning in an easily extensible way. 
In particular, {\tool} allows the interactive selection of subpopulations guided by a topological summary of the data and applies data mining and machine learning to these selected subpopulations for in-depth exploration.

\end{abstract}


\section{Introduction}
\label{sec:introduction}

High-throughput technologies have made field observations possible at scale in numerous branches of life sciences. In medicine, easy and increased access to imaging technologies and assay instruments have made it possible to capture a patient's biomedical trajectory in hospitals as the patient responds to various drugs and therapies. 
Similarly, in agricultural biotechnology, crop phenotyping technologies and on-field sensors are being widely adopted to capture a whole array of plant phenotypic traits (both morphological and physical) and growth environments (weather and soil parameters). Consequently, there has been an explosion of a new type of complex, multidimensional data called \emph{phenomics} data  \cite{houle2010phenomics,zhao2019crop}, which is a combination of phenotypic and environmental data and genotypic data collected using sequencing technologies.  Phenomics is the branch of modern biology that deals with the analysis of phenomics data to elucidate the science behind genotype ($G$) to environment ($E$) interactions toward controlling phenotypic ($P$) performance, that is, the study of the mapping $G\times E\rightarrow P$ \cite{houle2010phenomics,tardieu2017plant, zhao2019crop}. 

The use of phenomics has been exemplified in plant and agricultural biotechnology. Consider a commodity crop such as corn grown in various regions across the continental United States, Canada, and Mexico. With hundreds of varieties (genotypes) grown over thousands of geographical locations, the effects of growth environment on different genotypes can be significant. More specifically, the same genotype could show varying phenotypic trait performances in different environments (plasticity) \cite{kusmec2018harnessing}, or different genotypes may respond to environmental stresses differently \cite{cairns2012maize}. Under these circumstances, conducting only genotype to phenotype association studies is inadequate \cite{wray2013pitfalls}. Formulation of hypotheses on phenotypic control should involve the role of environment as well \cite{xu2016envirotyping}. Consequently, projects such as the Genomes-to-Field initiative 
\cite{alkhalifah2018maize, lawrence2019idea} have initiated a community-wide call to generate large-scale phenomics data collections. These projects are collecting vast swathes of phenomic data spanning hundreds of genotypes grown in diverse geographical regions, and tracking multiple phenotypic traits and their growth environments. 

Analyzing these large-scale high-dimensional datasets can be challenging.  Firstly, these datasets are generated without any particular central driving hypothesis. 
In fact, the role of bioinformaticians is to use data analysis to extract data-guided hypotheses. 
Secondly, the more traditional Genome Wide Association Studies (GWAS) \cite{bush2012genome} tools are not readily suited to handle environmental data. 
The multitude of dimensions poses a more severe challenge to more traditional multivariate statistical methods. Whereas the recourse is to use dimensionality reduction tools such as principal component analysis (PCA), these tools typically work well only to elucidate coarse population-level correlations or dominant dimensions. 
However, what is of interest to this community are the \emph{subpopulation-level variabilities}, e.g.,~certain environmental factors tend to better control a specific trait in a certain \emph{subset} of genotypes (subpopulations) than in others. 
Furthermore, the lack of effective large-scale data visualization tools poses additional challenges to the domain scientist trying to explore the data. 

In this paper, we present a software tool called {\tool} that is designed to overcome the above challenges associated with phenomics data analysis. 
{\tool} is a domain-specific adaptation of a toolbox called the \textsf{Mapper Interactive}~\cite{ZhouChalapathiRathore2021} and specifically targets the analysis and visualization of multidimensional phenomics data. 
More specifically, in comparison to existing approaches, the main advantage of {\tool} is two-fold. 
First, it provides rich, interactive capabilities in the exploratory analysis of phenomics data. In particular, users can select subpopulations of the data guided by their graph-based, multiscale topological summaries for in-depth analysis. 
Second, as a property inherent from the design of \textsf{Mapper Interactive}, {\tool} integrates visual analytics with data analysis and machine learning module in an easily extensible way, thus supporting the exploration of selected subpopulations of phenomics data with built-in and new analysis and visualization modules. 
Finally, {\tool} is open source, available at \url{https://github.com/tdavislab/PhenoMapper}.

\section{Related Work}
\label{sec:related}

We review the state-of-the-art computational tools for phenomics data analysis and visualization, with a focus on topological techniques. 

\para{Phenomics data analysis.}
Phenomics is a relatively nascent field. 
Most of the current automated tools focus on phenotyping, i.e.,  phenotypic data gathering, which predominantly involves image analysis to extract spatiotemporal features of interest on the field \cite{zhao2019crop}. 
Analysis of genotypic data in relation to their role in controlling a phenotype is typically conducted through Genome Wide Association Studies (GWAS) \cite{bush2012genome} tools, which aim to identify correlations between the genomic variations (i.e., Single Nucleotide Polymorphisms or SNPs) and phenotypic trait values. However, environmental variability is not captured with these tools. 

\para{Topological data analysis for phenomics.}
Topological data analysis, and more specifically the \emph{mapper} framework proposed by Singh {\etal}~\cite{SinghMemoliCarlsson2007, LumSinghLehman2013}, offers a scalable approach to exploration of the multidimensional phenomic datasets.
The coordinate-free representations supported by the mapper framework alongside other desirable features such as its robustness to noise and its readiness to visualization \cite{LumSinghLehman2013} make the framework suited for phenomics data analysis. 
In a recent work, Kamruzzaman {\etal}~\cite{KamruzzamanKalyanaramanKrishnamoorthy2019} developed \textsf{Hyppo-X}, an open-source implementation of the mapper  framework.  Given a large phenomics data collection as a $d$-dimensional point cloud, \textsf{Hyppo-X} first generates topological objects (1D-skeletons of simplicial complexes, referred to as \emph{mapper graphs}) of the data based on a user-specified selection of filter functions (i.e., environmental dimensions) and phenotypic variable(s) of interest.
Furthermore, the tool supports visualization of this mapper graph and extraction of several structural features such as paths \cite{KalyanaramanKamruzzamanKrishnamoorthy2019} and flares \cite{kamruzzaman2018detecting}. 
Domain scientists use \textsf{Hyppo-X} to combine the visual representations and extracted features to formulate specific hypotheses.
The \textsf{Hyppo-X} tool has demonstrated the utility of topological data analysis on both plant phenomics data \cite{KamruzzamanKalyanaramanKrishnamoorthy2019} and biomedical patient trajectories \cite{madhobi2019visual}, but the tool lacks interactivity and downstream machine learning (ML) capabilities -- both important from the perspective of domain scientists trying to navigate and explore the data for hypothesis discovery.

Table~\ref{table:compare} presents a summary of the functional differences between \textsf{Hyppo-X}~\cite{KamruzzamanKalyanaramanKrishnamoorthy2019} and {\tool}. 
The main differences are that (a) {\tool} offers more interactive capabilities in exploring phenomics data via its mapper graph representation, including easily extensible GUI, and (b) it provides easily extensible data analysis and ML capabilities including regression, feature selection, and dimensionality reduction. 
As a domain-specific adaptation of \textsf{Mapper Interactive}~\cite{ZhouChalapathiRathore2021}, {\tool} inherits a number of properties including extendability and interactivity. 
With these additional capabilities, {\tool} is well suited for the analysis and visualization of phenomics data, in particular, for exploring the subpopulations. 
\begin{center}
\begin{table}[h]
\centering
\begin{tabularx}{0.75\columnwidth}{|X|m{0.03\textwidth}|m{0.03\textwidth}|} 
\hline
Features & HX & PM \\ \hline
\rowcolor{aliceblue}\multicolumn{3}{|c|}{Mapper graph computation and visualization} \\ \hline
Mapper graph layout adjustment  &  \cellcolor{babypink}{N} & \cellcolor{babyblue}{Y} \\ \hline
Interactive parameter adjustment  &  \cellcolor{babyblue}{Y} & \cellcolor{babyblue}{Y} \\ \hline
On-the-fly mapper graph computation &  \cellcolor{babyblue}{Y} & \cellcolor{babyblue}{Y} \\ \hline
\rowcolor{aliceblue}\multicolumn{3}{|c|}{Interactive visualization} \\ \hline
Node selection  &  \cellcolor{babyblue}{Y} & \cellcolor{babyblue}{Y} \\ \hline
Path selection  &  \cellcolor{babyblue}{Y} & \cellcolor{babyblue}{Y} \\ \hline
Subgraph (connected component/cluster) selection  &  \cellcolor{babypink}{N} & \cellcolor{babyblue}{Y} \\ \hline
Structural feature extraction (flares and special paths) &  \cellcolor{babyblue}{Y} & \cellcolor{babypink}{N} \\ \hline 
Easily extensible GUI & \cellcolor{babypink}{N} & \cellcolor{babyblue}{Y} \\ \hline
\rowcolor{aliceblue}\multicolumn{3}{|c|}{Data analysis and ML modules} \\ \hline
Semi-automatic hypothesis formulation & \cellcolor{babyblue}{Y} &  \cellcolor{babyblue}{Y} \\ \hline
Easily extensible analysis and ML modules & \cellcolor{babypink}{N} & \cellcolor{babyblue}{Y} \\ \hline
Regression & \cellcolor{babypink}{N} & \cellcolor{babyblue}{Y} \\ \hline
Feature selection & \cellcolor{babypink}{N} & \cellcolor{babyblue}{Y} \\ \hline
Dimensionality reduction & \cellcolor{babypink}{N} & \cellcolor{babyblue}{Y} \\ \hline
\rowcolor{aliceblue}\multicolumn{3}{|c|}{Controls and other features} \\ \hline
Import and export selected subpopulations  &  \cellcolor{babypink}{N} & \cellcolor{babyblue}{Y} \\ \hline
Open source implementation & \cellcolor{babyblue}{Y}  & \cellcolor{babyblue}{Y} \\ \hline
\end{tabularx}
\vspace{1mm}
\caption{Comparing features of \textsf{Hippo-X} (HX) against {\tool}.  Blue means ``yes'' (Y), pink means ``no'' (N).}
\label{table:compare}
\vspace{-6mm}
\end{table}
\end{center}

\section{Technical Background}
\label{sec:background}

{\tool} is created with the mapper framework at its core~\cite{SinghMemoliCarlsson2007}. 
It visualizes the 1D skeleton of a mapper construction, called the \emph{mapper graph}, for a phenomics dataset given as a point cloud.

\begin{figure}[!h]
 \centering
 \includegraphics[width=0.6\columnwidth]{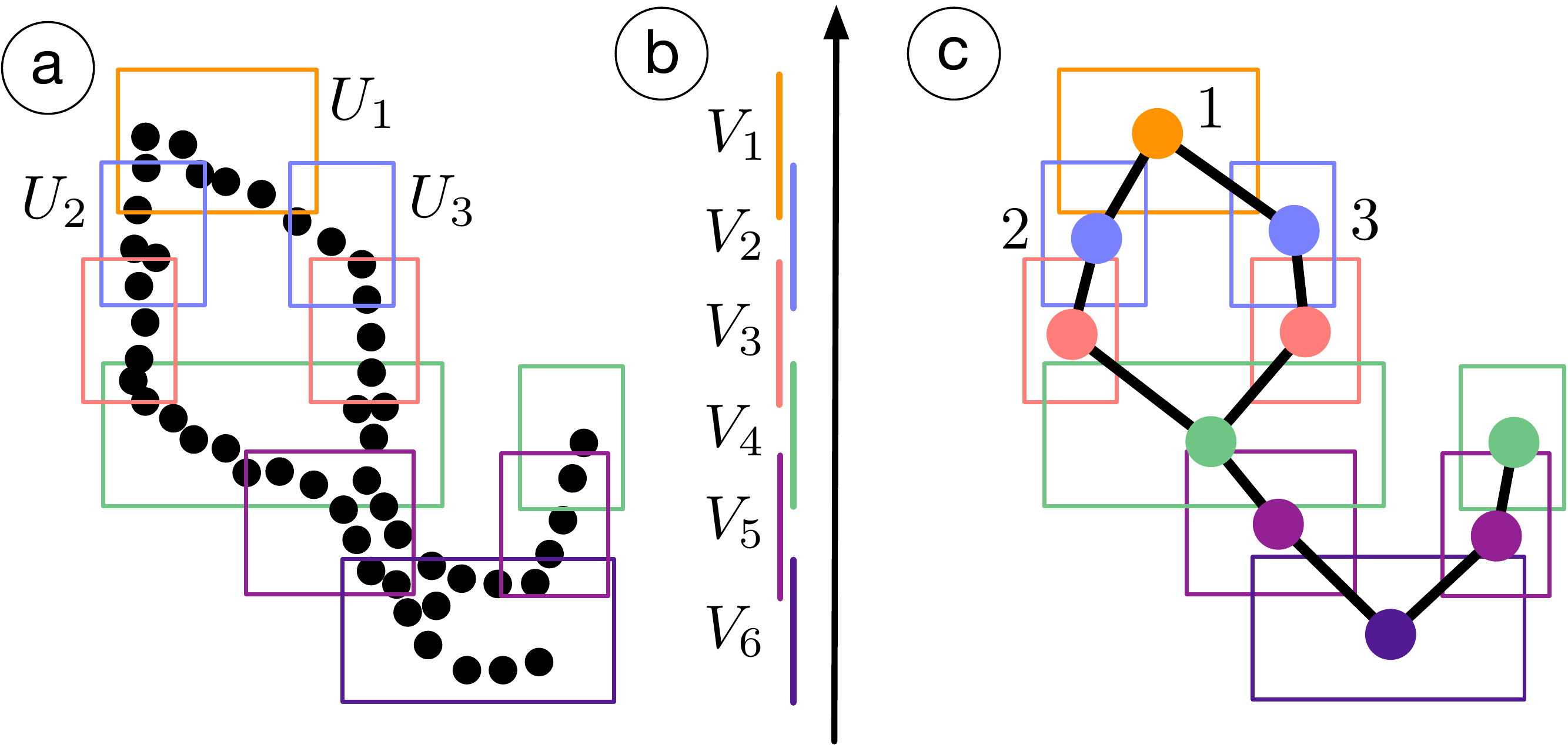}
 \vspace{-6mm}
 \caption{A mapper graph of a point cloud sampled from the silhouette of a kite. See text for more details.}
 \label{fig:kite}
  \vspace{-2mm}
\end{figure}

To define the mapper graph, we first explain the nerve of a cover.   
Let $\Xspace \subset \Rspace^d$ (for $d \geq 1$) denote a potentially multidimensional point cloud. 
A \emph{cover} of $\Xspace$ is a set of open sets in $\Rspace^d$, $\Ucal = \{U_i\}^{l}_{i=1}$ such that $\Xspace \subset \cup_{i} U_i$. 
The 1D nerve of $\Ucal$ is a graph and is denoted as $\Ncal_1(\Ucal)$. 
Each node $i$ in $\Ncal_1(\Ucal)$ represents a cover element $U_i$, and an edge exists between two nodes $i$ and $j$ if $U_i \cap U_j$ is nonempty for the corresponding cover elements. 
\autoref{fig:kite}a gives an example in which $\Xspace$ is a 2D point cloud sampled from the silhouette of a kite. 
The cover $\Ucal$ of $\Xspace$ consists of a collection of 10 rectangles on the plane. 
The 1D nerve of $\Ucal$ is the graph in~\autoref{fig:kite}c.  

In the original mapper construction introduced by Singh \etal~\cite{SinghMemoliCarlsson2007}, obtaining a cover is guided by a set of scalar functions defined on $\Xspace$. 
For simplicity, we define a mapper graph with a single scalar function $f: \Xspace \to \Rspace$.  
We start with a finite cover of $f(\Xspace)$ using intervals, that is, a cover $\Vcal = \{V_k\}_{k=1}^{n}$ of $f(\Xspace) \subset \Rspace$ such that $f(\Xspace) \subseteq \cup_{k} V_k$,  see~\autoref{fig:kite}b. 
We obtain a cover $\Ucal$ of $\Xspace$ by considering the clusters induced by points in $f^{-1}(V_k)$ for each $V_k$ as a cover element.
The 1D \emph{mapper graph} of $(\Xspace, f)$, denoted as $\Mcal$, is the 1D nerve of $\Ucal$, $\Mcal:= \Ncal_1(\Ucal)$.

We use \autoref{fig:kite} as an example where the point cloud $\Xspace$ is equipped with a height function $f: \Xspace \to \Rspace$. 
A cover $\Vcal = \{V_1, \cdots, V_6\}$ of $f(\Xspace)$ is formed by six intervals. 
For each $k$ ($1 \leq k \leq 6$), $f^{-1}(V_k)$ induces a number of  clusters that are subsets of $\Xspace$. 
Such clusters form elements of a cover of $\Xspace$.
For instance, as illustrated in \autoref{fig:kite}a, $f^{-1}(V_1)$ induces a single cluster of points that is enclosed by the orange cover element $U_1$, and $f^{-1}(V_2)$ induces two clusters of points enclosed by the blue cover elements $U_2$ and $U_3$. 
The mapper graph in \autoref{fig:kite}c has an edge between nodes  $1$ and $2$ since $U_1 \cap U_2 \neq \emptyset$. 
Note how the mapper graph captures the overall shape of the kite. 

Given a point cloud $\Xspace$, several parameters are needed to compute the mapper graph $\mapper$, including a function $f: \Xspace \to \Rspace$, referred to as the \emph{filter function}, the number of cover elements $n$ and their percentage of overlaps $p$, the metric $d_\Xspace$ on $\Xspace$, and the clustering method. 
For example, in \autoref{fig:kite}, $f$ is the height function, $n=6$ and $p=30\%$, $d_\Xspace$ is the Euclidean distance, and the clustering method is DBSCAN~\cite{EsterKriegelSander1996}. 

In practice, the choice of the filter function $f$ is nontrivial. 
Typically, a different choice of $f$ gives rise to a different type of summary.
Common choices for $d_\Xspace$ include the $L_2$-norm, variants of geodesic distances, and eccentricity~\cite{BiasottiGiorgiSpagnuolo2008, SinghMemoliCarlsson2007}. 
A common choice for the clustering method is DBSCAN~\cite{EsterKriegelSander1996}, which is a density-based clustering algorithm. 
DBSCAN has two parameters: $\epsilon$ is the neighborhood size of a given point, and $minPts$ is the minimum number of points needed to consider a collection of points as a cluster.

The filter function $f$ may be generalized to a multivariate function, that is, $f: \Xspace \to \Rspace^m$ (for $m \geq 2$). 
In most practical scenarios, $m = 2$, and the resulting mapper graph is referred to as a \emph{2D mapper graph}. The corresponding cover elements of $\Rspace^2$ become rectangles.
$\tool$ supports the computation of both 1D and 2D mapper graphs.

\section{Design and Implementation}

{\tool} is a domain-specific adaptation of \textsf{Mapper Interactive}~\cite{ZhouChalapathiRathore2021}, which is an interactive, extendable, and scalable toolbox for exploring generic, high-dimensional point clouds. We extend the original toolbox of \emph{Mapper Interactive} by adding new capabilities that are desirable for studying phenomics datasets.

\begin{figure}[!ht]
\centering
\includegraphics[width=0.8\columnwidth]{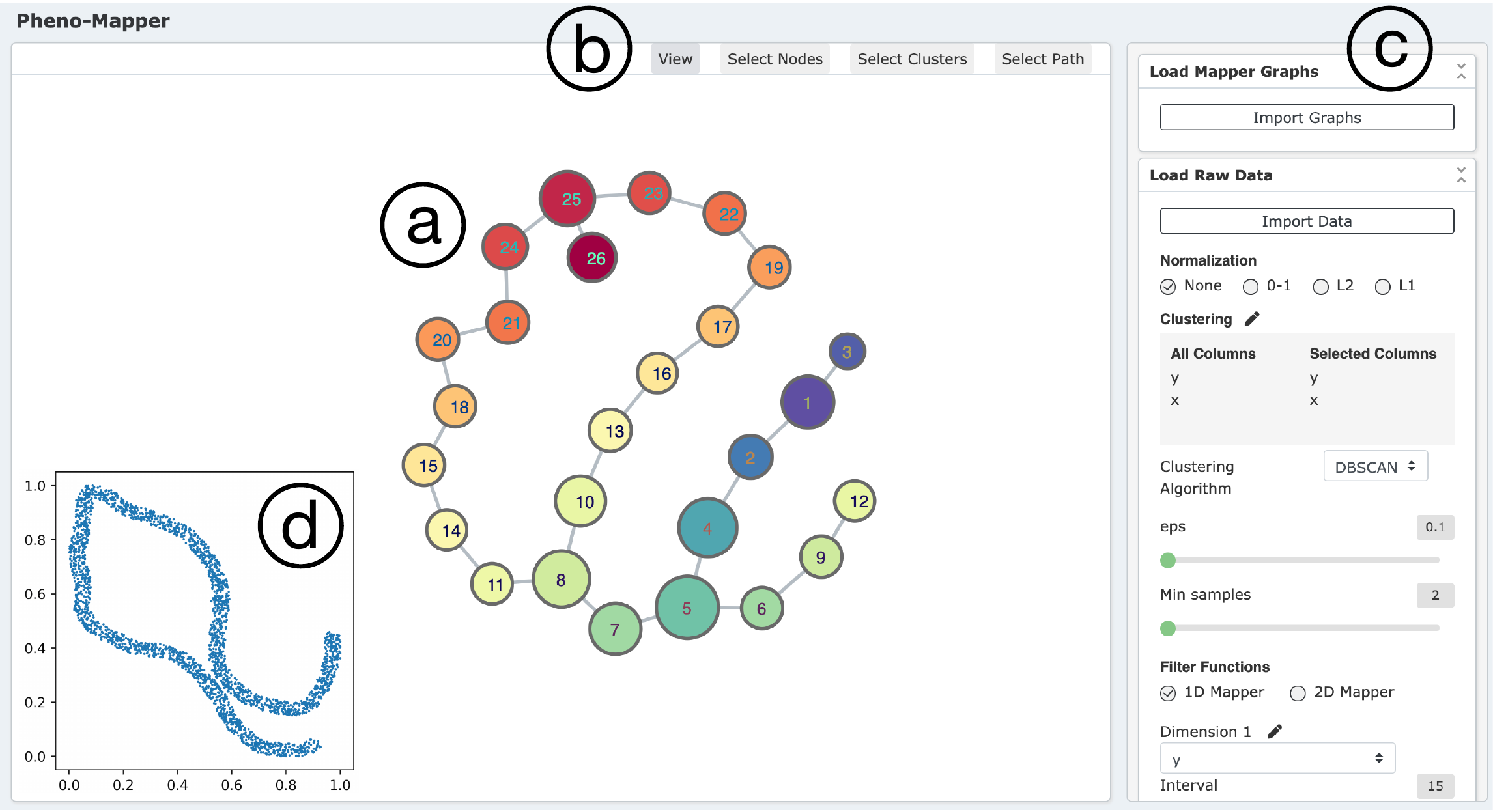}
\vspace{-2mm}
\caption{The user interface of {\tool}.}
\label{fig:interface}
\vspace{-2mm}
\end{figure}

The user interface of {\tool} is shown in~\autoref{fig:interface}. The interface consists of three panels. The \textbf{graph visualization panel} (a) visualizes a resulting mapper graph. The \textbf{selection panel} (b) enables three ways to select groups of nodes (\emph{subpopulations}), including the selection of individual nodes,  connected components (clusters), and nodes connected along a path. 
The \textbf{control panel} (c) provides various parameter controls for computing a mapper graph. Here, (a) displays a mapper graph computed from the point cloud of the silhouette of a kite~(d) from~\autoref{fig:kite}.

Compared with \textsf{Mapper Interactive}, {\tool} has the following new features to support in-depth exploration of phenomics data, with a specific focus on the analysis and visualization of subpopulations. 

\begin{figure}[!ht]
\centering
\includegraphics[width=0.75\columnwidth]{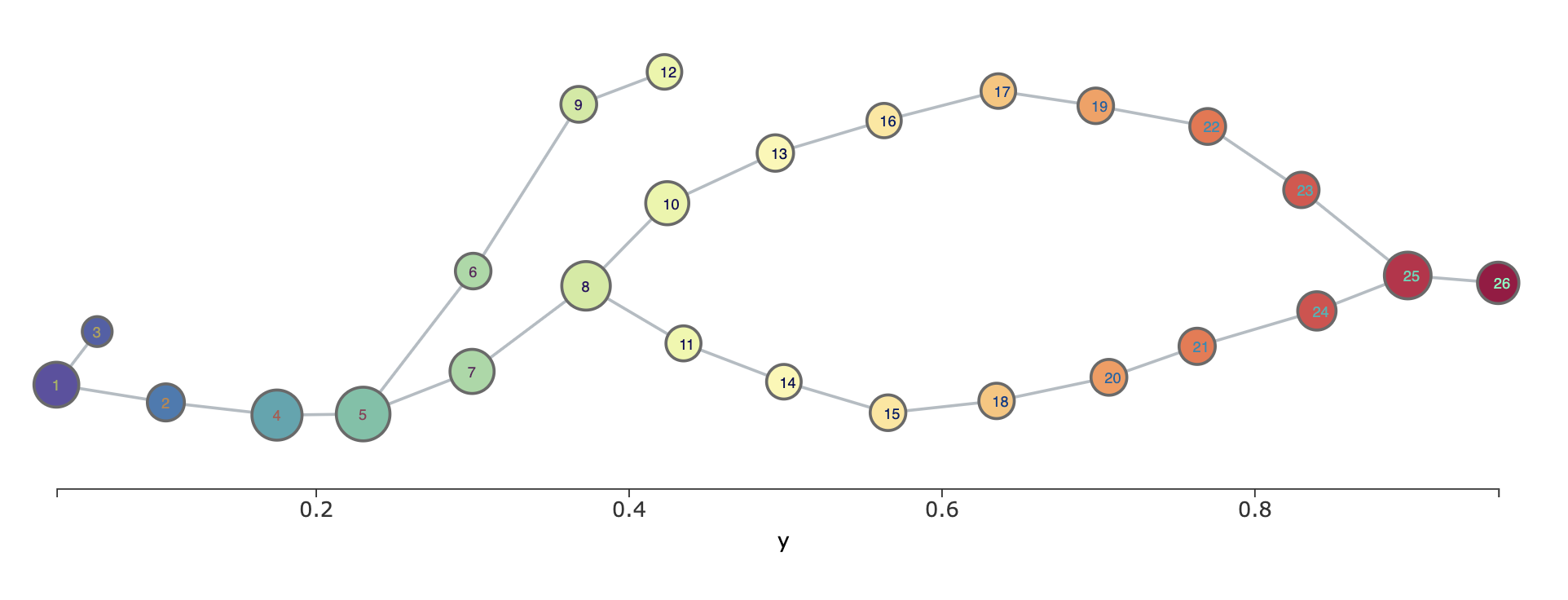}
\vspace{-8mm}
\caption{An alternative layout of the mapper graph in~\autoref{fig:interface} by aligning the nodes horizontally along a filter function.}
\label{fig:kite-sorted}
\end{figure}

\para{Alternative mapper graph layouts.} 
The default layout of a mapper graph is a force-directed layout~\cite{Kobourov2013}. 
Force-directed layouts are a class of graph drawing algorithms that assign forces among the edges and nodes of a graph and minimize the energy associated with these forces to achieve an aesthetically pleasing graph visualization. 
To study plant phenomics data with {\tool}, we provide an alternative layout option, which aligns the nodes of the graph along with a filter function that increases in the x-axis direction. 
For a 2D mapper graph, users are able to select one of the two filter functions for aligning the nodes. An example of this layout is shown in~\autoref{fig:kite-sorted}. 
Under this layout, nodes can be adjusted vertically to produce a more readable mapper graph w.r.t. the changes of a chosen filter function. 
In the setting of multidimensional plant phenomics datasets (see~\autoref{sec:ksne} and \autoref{sec:irrigation}), variables/dimensions such as time or days after planting (\textbf{DAP}) are important factors that characterize the growth of a plant; thus such an alternative mapper graph layout is particularly useful to emphasize the organizational principle of the population w.r.t. such variables. 

We provide a number of new modules for in-depth analysis of mapper graphs, using feature selection, scatter plots, and nonlinear dimensionality reduction (t-SNE).

\para{Feature selection.}
Since phenomics datasets usually contain information related to categorical variables such as genotypes, we provide a feature selection module, which can be used to determine important features for classification tasks related to various genotypes. 
Feature selection is performed using a linear support vector classification model (e.g., SVM), and the selected features can then be included to construct mapper graphs and to help separate nodes into different subpopulations based on the targeting genotypes for downstream analysis and visualization.

\para{Scatter plots.} 
We provide a scatter plot module to highlight correlations between a pair of dimensions, where points can be colored with any numerical or categorical dimension. 
Using scatter plots, for instance, users are able to identify the relation between two environmental variables and determine whether to include these variables within a mapper graph construction process. 
Furthermore, scatter plots can also be used to further investigate and confirm certain hypothesis generated by the mapper graph or another analysis module. 

\para{Dimensionality reduction.}
The t-SNE module is used as a supplement to the existing PCA module. 
Using either linear (e.g., PCA) or nonlinear (e.g., t-SNE) dimensionality reduction with selected subpopulations from a mapper graph, users will have a better understanding of the nature of the data.

\para{Export and import selected subpopulations.}
With {\tool}, users can export either the entire point cloud associated with a mapper graph (the entire population) or selected subgraphs of a mapper graph (subpopulations) for analysis. 
The exported population or subpopulations are saved into a JSON file, which contains the information of nodes, node relations (edges), and node cluster memberships. 
Such information can be used for further analysis, such as comparisons between different subpopulations.  

\para{Implementation details.}
{\tool} -- as an extension of \textsf{Mapper Interactive}~\cite{ZhouChalapathiRathore2021} -- is implemented using the standard HTML, CSS, Javascript stack with \emph{D3.js}, and \emph{JQuery} libraries. It is equipped with a Python backend using a \emph{Flask}-based server. The mapper graph computation is an accelerated version of \emph{KeplerMapper}~\cite{VeenSaulEargle2019} implementation. 
The ML modules (linear regression, feature selection, etc.) interface with Python libraries \emph{scikit-learn} and \emph{statsmodels}; such a design makes {\tool} easily extendible to include other ML modules available from \emph{scikit-learn} with only a few lines of code.

\section{Use Case: KS/NE Dataset}
\label{sec:ksne}

With {\tool}, users can summarize and interactively interpret phenomics datasets under the mapper framework. 
In this and the next section, we showcase the analysis and visualization capabilities of {\tool} using two real-world plant (e.g., maize) phenomics datasets  first studied by Kamruzzaman \etal~\cite{KamruzzamanKalyanaramanKrishnamoorthy2019}. 
A portion of the findings in these use cases can be obtained from existing frameworks (e.g.,~\textsf{Hyppo-X}~\cite{KamruzzamanKalyanaramanKrishnamoorthy2019}), but the main advantage of {\tool} is that (a) it provides interactive explorations of phenomics data, and (b) it integrates visual analytics with machine learning in an easily extensible way.   
{\tool} not only provides insights into variabilities across different subpopulations of the maize datasets across multiple scales, but also supports in-depth analysis of such subpopulations with machine learning techniques such as feature selection and regression. 

The first maize dataset, referred to as the \textbf{KS/NE} dataset~\cite{KamruzzamanKalyanaramanKrishnamoorthy2019},   contains the growth information of two maize genotypes (type A and type B) that were cultivated in Kansas (KS) and Nebraska (NE) in the United States. 
It consists of 400 rows (data points) describing the phenotypic and environmental measurements for a number of maize plants. 
It describes daily measurements of maize plants across the first 100 days of the growing season for each of the four (location, genotype) combinations: (KS, A), (KS, B), (NE, A), and (NE, B).   
The columns consist of the genotype of each plant (\textbf{A} or \textbf{B}), a time measurement recording the days after planting (\textbf{DAP}), the \textbf{growth rate} of each plant, and 10 environmental variables such as \textbf{humidity}, \textbf{temperature}, \textbf{rainfall}, \textbf{solar radiation}, \textbf{soil moisture}, and \textbf{soil temperature}. 

\subsection{Reproducing Known Results}
To first reproduce the experimental result of Kamruzzaman \etal~\cite{KamruzzamanKalyanaramanKrishnamoorthy2019}, we  use the \textbf{growth rate} to form our point cloud and \textbf{DAP} as the filter function to construct a 1D mapper graph. 

\begin{figure}[!ht]
\centering
\includegraphics[width=0.8\columnwidth]{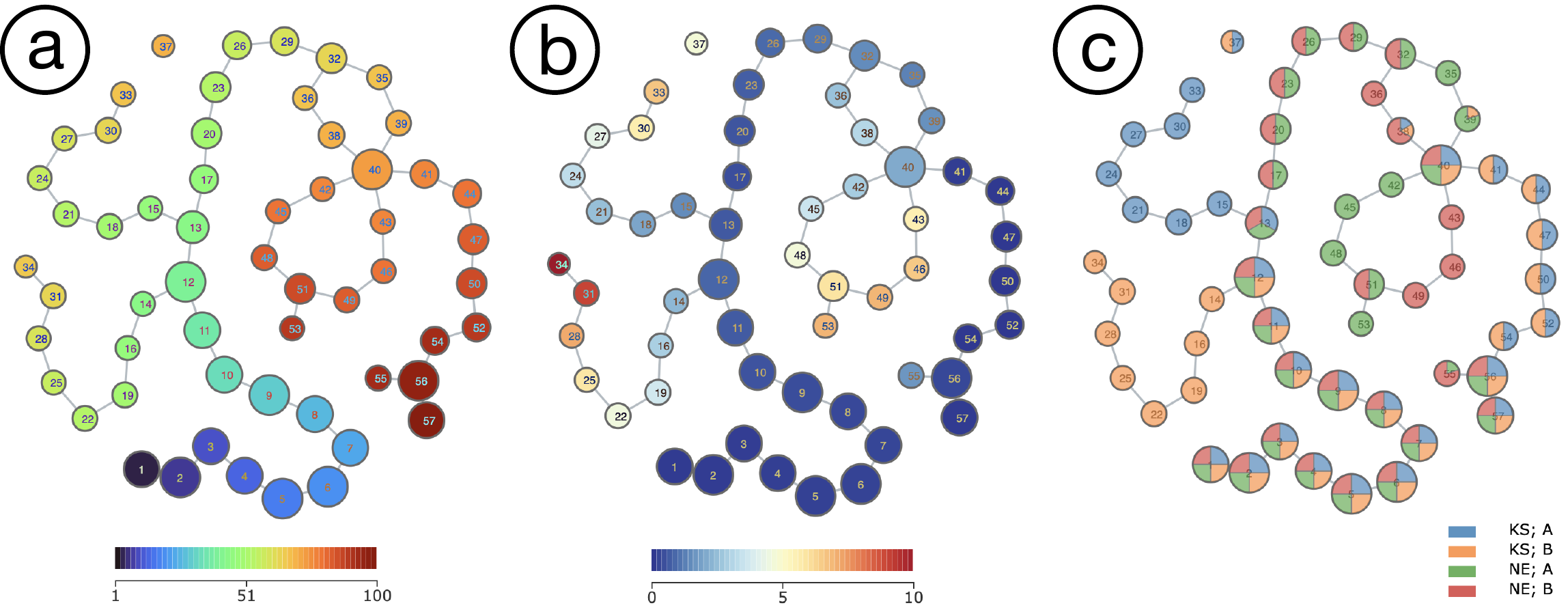}
\vspace{-2mm}
\caption{Mapper graphs of the KS/NE dataset: (a) nodes are colored by DAP; (b) nodes are colored by growth rate; (c) nodes are visualized with pie charts according to four (location, phenotype) categories. For DBSCAN, we chose $\epsilon = 0.6$,  $minPts = 2$. For mapper graphs, we set $n=30$, $p=25\%$.}
\label{fig:ksne}
\end{figure}

The resulting mapper graphs are shown in~\autoref{fig:ksne}.
In~\autoref{fig:ksne}a, the node color represents the average \textbf{DAP} of points contained within the node. 
In~\autoref{fig:ksne}b, the node color represents the average \textbf{growth rate} of points contained within the node. 
In~\autoref{fig:ksne}c, the pie chart on each node represents the variations among (location, phenotype) combinations for plants contained in the node. 
The node size represents the number of points contained in each node. 
 
\begin{figure}[!ht]
\centering
\includegraphics[width=0.55\columnwidth]{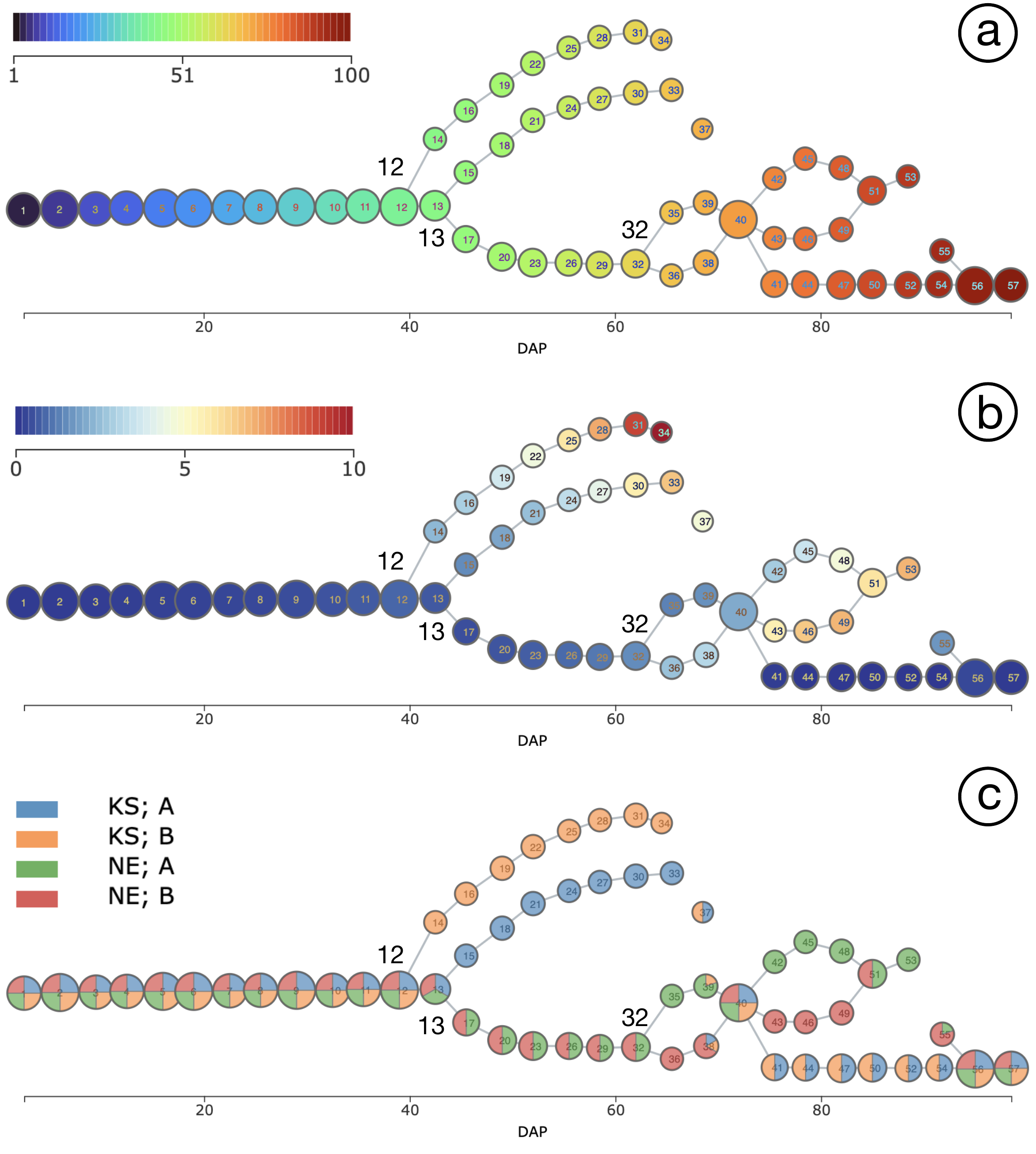}
\vspace{-4mm}
\caption{Mapper graphs of the KS/NE dataset aligned along the horizontal time axis quantified by DAP: (a) nodes are colored by DAP; (b) nodes are colored by growth rate; (c) nodes are visualized with pie charts according to four (location, phenotype) categories.}
\label{fig:ksne-time}
\end{figure}

To highlight the time axis, we use an alternative mapper graph layout of~\autoref{fig:ksne} by aligning a chosen dimension (\textbf{DAP}) along the $x$-axis, as shown in~\autoref{fig:ksne-time}.  

We observe from~\autoref{fig:ksne-time}b that for the first few days (nodes 1 to 11), all plants have similar \textbf{growth rates}. 
Starting from node 12 (${\sim}$40 days), type B plants grown in KS, i.e.,  the orange (KS; B) category in~\autoref{fig:ksne-time}c, bifurcate from the main branch, and their \textbf{growth rates} start to accelerate faster than other plants (\autoref{fig:ksne-time}b). 
Starting from node 13 (${\sim}$43 days), type A plants grown in KS, i.e.,  the blue (KS, A) category, also bifurcate from the main branch with a faster \textbf{growth rate} than those grown in NE. 
The plants grown in NE from both genotypes $A$ and $B$ have a similar \textbf{growth rate} until node 32 (${\sim}$62 days) before bifurcating further into subpopulations. 
These results demonstrate that {\tool} helps the users quickly identify interesting subpopulations of plants that have different growth behaviors, and how such behaviors vary according to their phenotypes.
 
\subsection{Conducting New in-Depth Analysis}
The main advantage of {\tool} over existing tools such as \textsf{Hyppo-X} is that {\tool} integrates the visual selection of subpopulations with various data analysis and ML modules for in-depth analysis. 
In addition, {\tool} can be easily extended to include additional analysis modules with a few lines of code. 
By using {\tool} for the KS/NE dataset, we will perform in-depth analysis of the entire population as well as selected subpopulations using feature selection and regression. 
In particular, we will use the KS/NS dataset to highlight {\tool}'s analysis and visualization capabilities. 

\para{Linear regression of the entire population.}
From the mapper graphs in~\autoref{fig:ksne-time}, we can see that plants from the same genotype might have different phenotypes (e.g., \textbf{growth rate}) when they were grown in different locations. 
Since each data point contains dimensions describing the environmental information, we employ the linear regression module in {\tool} to further explore how the environmental factors affect the \textbf{growth rate}. 

\begin{figure}[!ht]
\centering
\includegraphics[width=0.8\columnwidth]{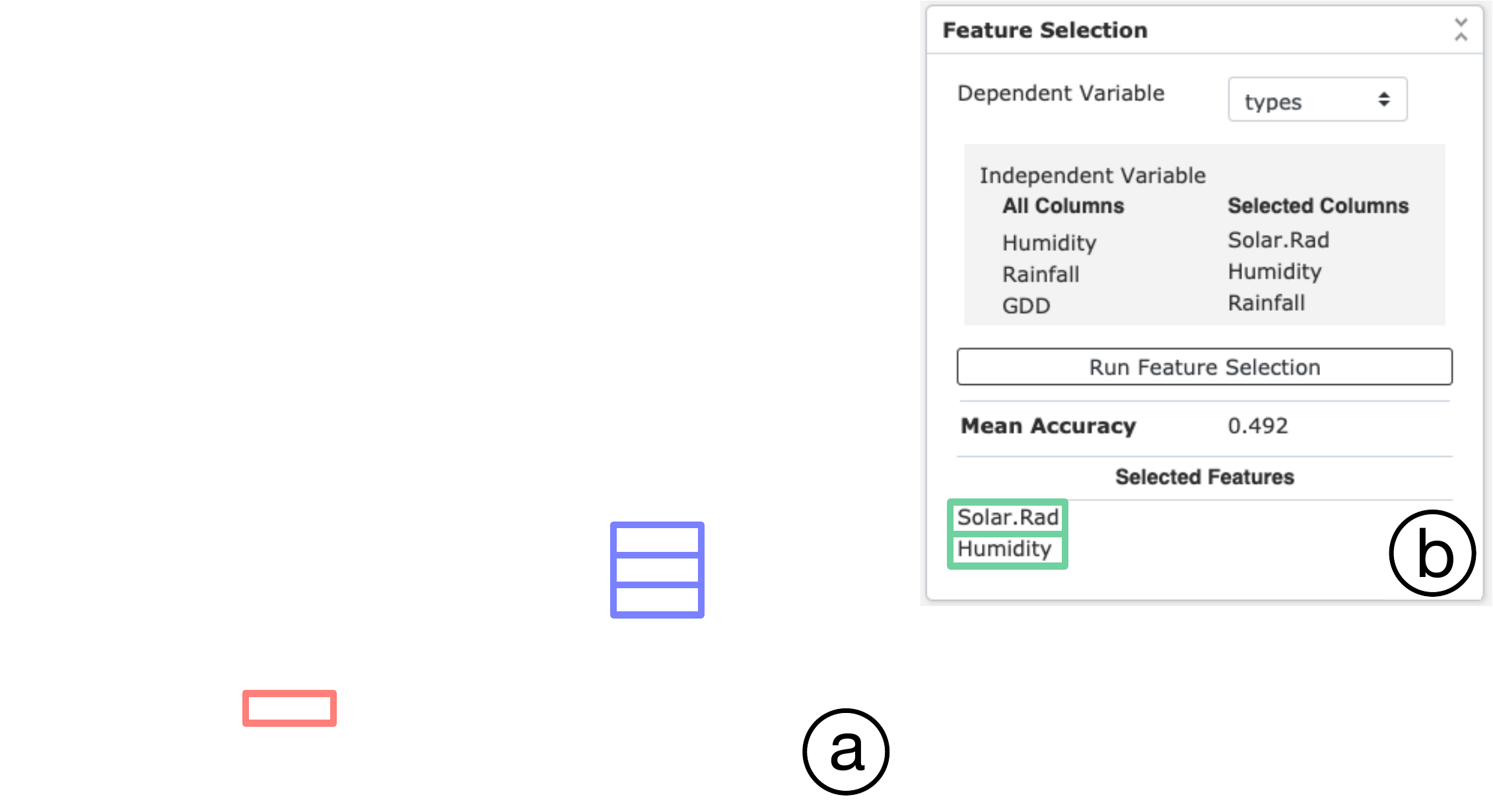}
\vspace{-2mm}
\caption{KS/NE dataset: (a) linear regression of environmental variables against the growth rate; (b) feature selection.}
\label{fig:ksne-regression}
\vspace{-2mm}
\end{figure}

The linear regression result is shown in~\autoref{fig:ksne-regression}a. The R-squared value ($0.119$, red box) is low, meaning that the proportion of variability explained by this model is low, and getting precise predictions from the model is difficult. 
However, under the significant level ($p$-value) of $0.05$, the variables \textbf{solar radiation}, \textbf{humidity}, and \textbf{rainfall} are significantly correlated with \textbf{growth rate} (blue boxes), which means the relation between these environmental variables and the growth rate is still statistically significant. 
This result indicates that the phenotypic behaviors of the plants are likely affected by these environmental factors. 

\para{Feature selection.}
To further explore how the environmental variables can help better separate different (location, genotype) combinations in the resulting mapper graph, we consider two ways to incorporate the environmental information to the mapper framework. 
The first way is to add one environmental variable as a filter function and construct 2D mapper graphs. 
The second way is to encode the environmental variables as additional dimensions within a multidimensional point cloud for analysis.

To determine which environmental variables to considered as filter functions, we have added a new feature selection module within {\tool} to select the best variables based on the linear support vector classification model (e.g., SVM). 
The feature selection result is shown in~\autoref{fig:ksne-regression}b. The selected environmental variables are \textbf{humidity} and \textbf{solar radiation} (green boxes).

\begin{figure}[!ht]
\centering
\vspace{-2mm}
\includegraphics[width=0.55\columnwidth]{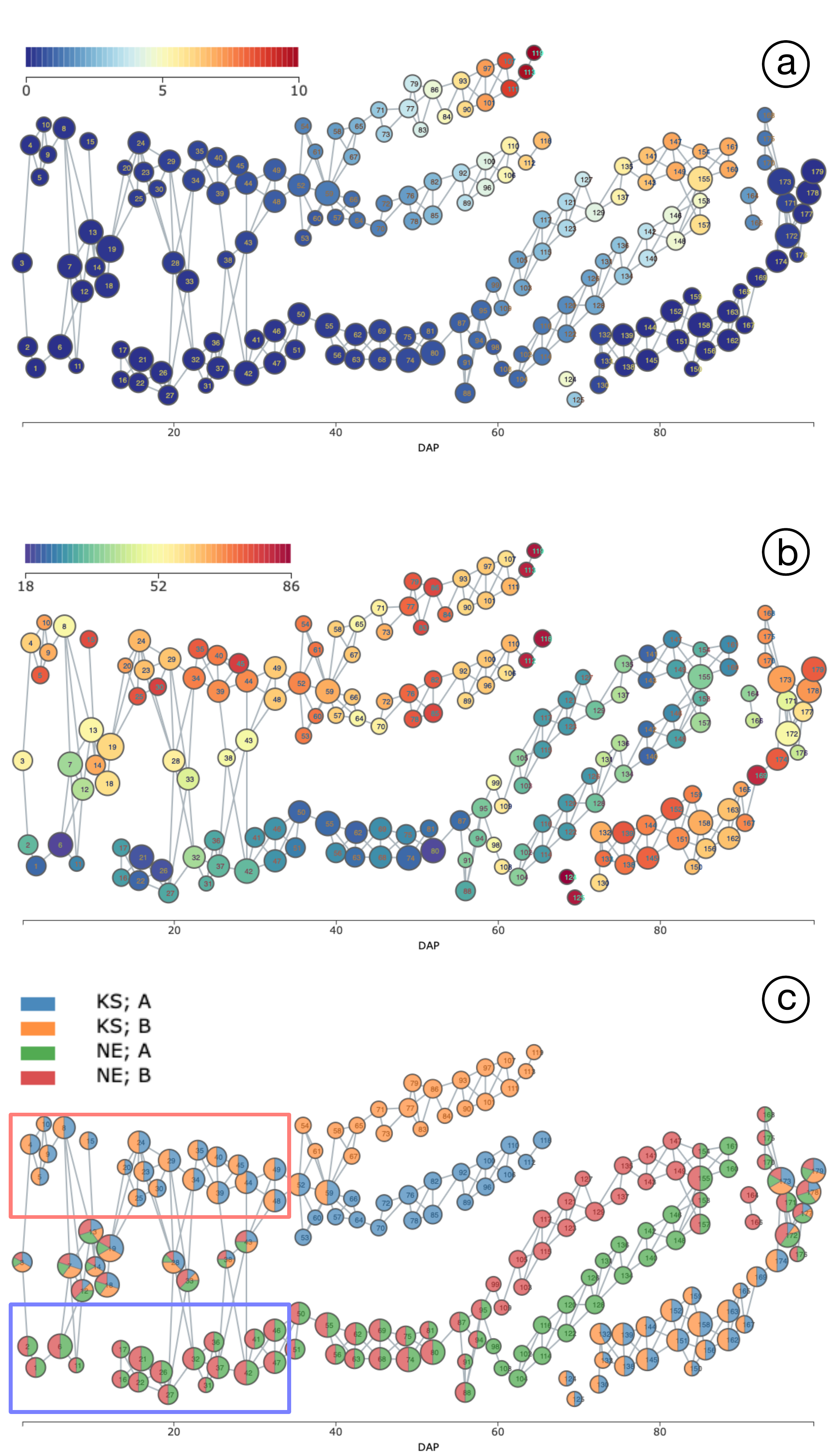}
\vspace{-2mm}
\caption{The 2D mapper graph of KS/NE data. (a) Nodes are colored by growth rate; (b) nodes are colored by humidity; (c) nodes are visualized with pie charts according to four (location, phenotype) categories. For DBSCAN, we chose $\epsilon = 0.6$,  $minPts = 2$. For mapper graph, we set $f_1=DAP$, $n_1=30$, $p_1=25\%$, $f_2=Humidity$, $n_2=5$, $p_2=50\%$. }
\label{fig:ksne-2d-time}
\end{figure}

\para{2D mapper graphs with humidity and DAP.}
We first add \textbf{humidity} as a second filter function in addition to \textbf{DAP} and construct a 2D mapper graph. 
The result is shown in~\autoref{fig:ksne-2d-time}. 
The 2D mapper graph retains a similar structure to the 1D mapper graph, but the plants are split earlier (w.r.t. to the DAP -- the x-axis) based on their planting locations. 
As shown in~\autoref{fig:ksne-2d-time}c, at the beginning of planting, a subpopulation of plants from KS (in the red box) bifurcates from other plants, and also a subpopulation of plants from NE (in the blue box) bifurcates as well. The result implies that the \textbf{humidity} variable provides additional information that helps split plants from different locations. In~\autoref{fig:ksne-2d-time}b, the nodes are colored with the average \textbf{humidity} values, which confirms that, in general, the average \textbf{humidity} values in KS are higher than those values in NE.

\begin{figure}[!ht]
\centering
\includegraphics[width=0.55\columnwidth]{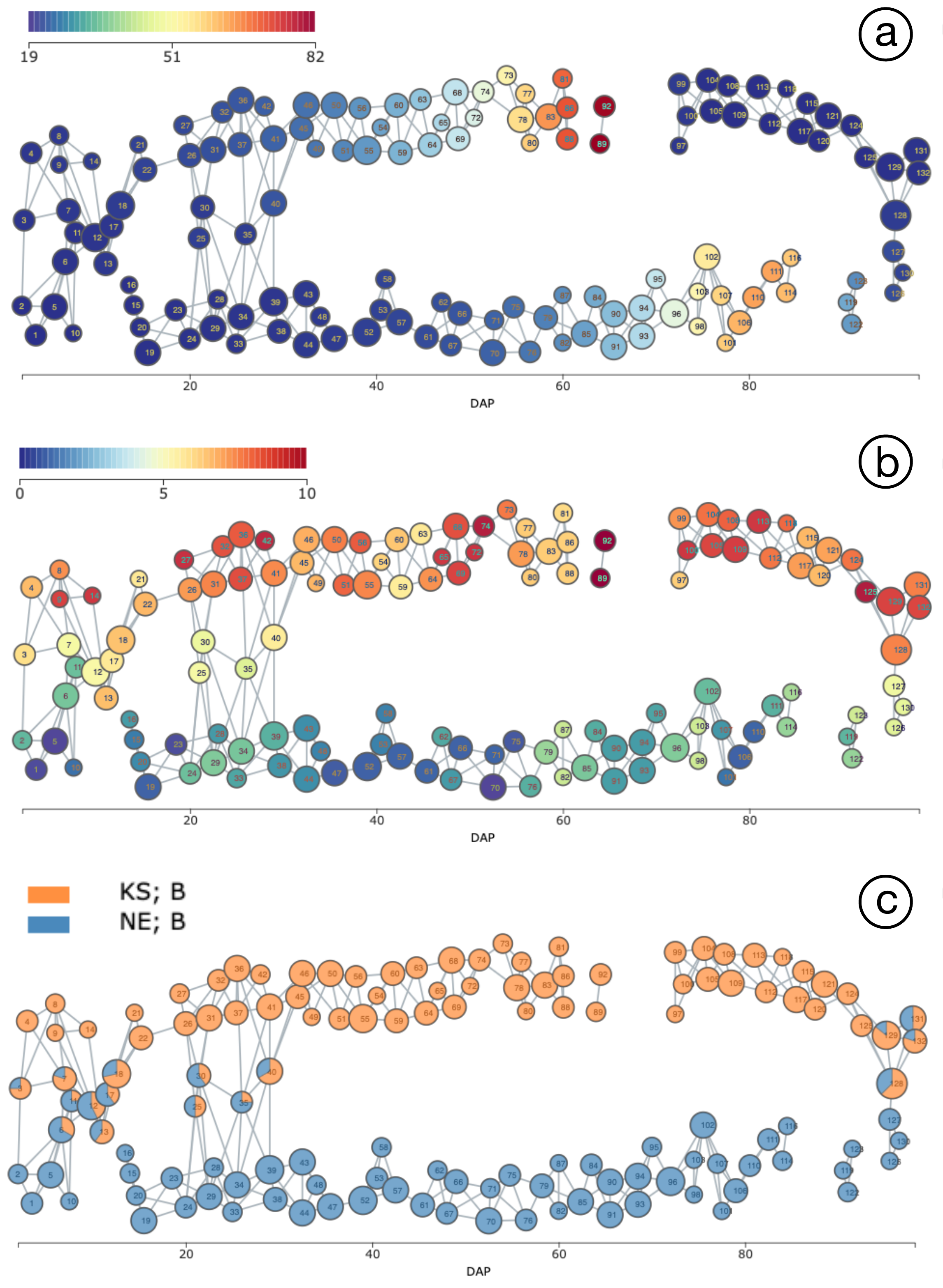}
\vspace{-4mm}
\caption{The 2D mapper graph of KS/NE data using only the plants from genotype B: (a) nodes are colored by growth rate; (b) nodes are colored by humidity; (c) nodes are shown with pie chars associated with the two categories (KS, B) and (KE, B). For DBSCAN, $\epsilon = 0.6$,  $minPts = 2$. For mapper graph, $f_1=DAP$, $n_1=30$, $p_1=46\%$, $f_2=Humidity$, $n_2=5$, $p_2=46\%$.}
\label{fig:ksne-B-2d-time}
\end{figure}

\para{Analyzing subpopulations with the same genotype.}
To better understand how humidity affects the growth rate of plants, we are interested in analyzing and comparing subpopulations of nodes from the same genotype but different locations. \autoref{fig:ksne-B-2d-time} demonstrates the 2D mapper graph constructed with plants from genotype B only at locations KS and NE using export/import function of {\tool}.  
In~\autoref{fig:ksne-B-2d-time}c, we see that using \textbf{DAP} and \textbf{humidity} as filter functions clearly separate the two categories: (KS, B) and (NE, B).  

\begin{figure}[!ht]
\centering
\includegraphics[width=0.55\columnwidth]{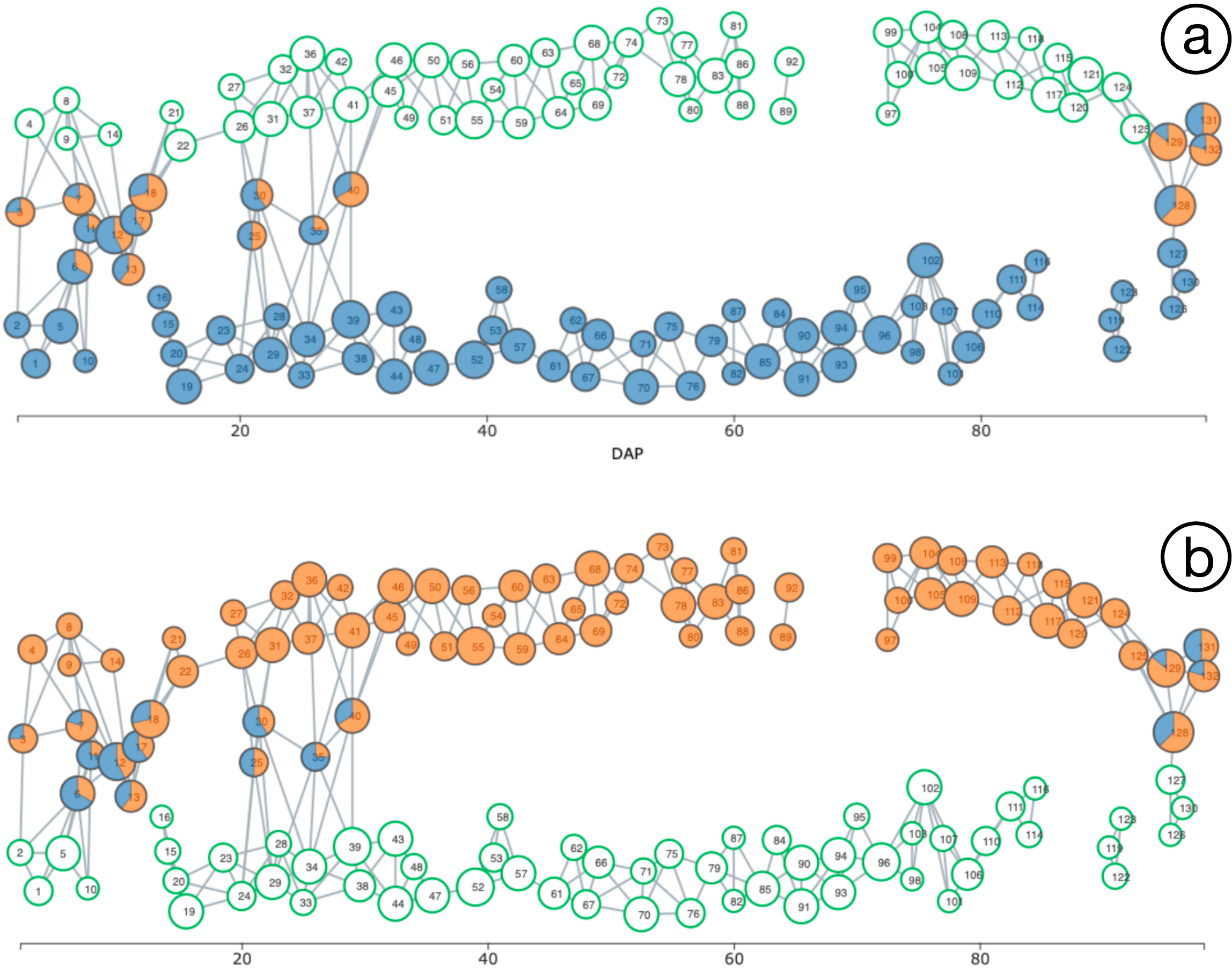}
\vspace{-4mm}
\caption{KS/NS dataset: green nodes in (a)-(b) correspond to the selected subpopulation for regression analysis.}
\label{fig:ksne-selection}
\end{figure}

We now perform linear regression on selected subpopulations, which are highlighted as green nodes in \autoref{fig:ksne-selection}a-b, respectively. 
Green nodes in \autoref{fig:ksne-selection}a form a subpopulation with plants from KS only, whereas those in \autoref{fig:ksne-selection}b are from NE only. 

\begin{figure}[!ht]
\centering
\includegraphics[width=0.8\columnwidth]{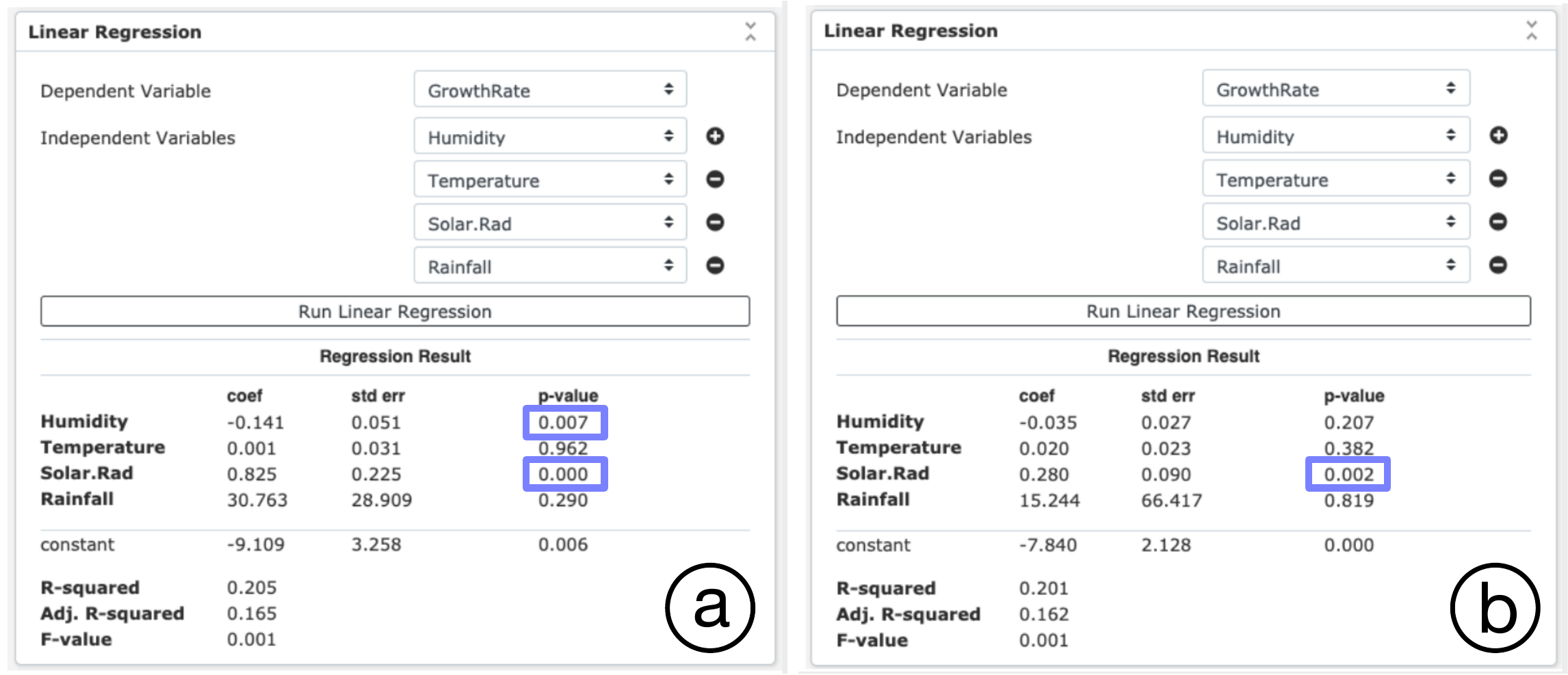}
\vspace{-4mm}
\caption{Linear regression results of the selected KS and NE subpopulations from~\autoref{fig:ksne-selection}a-b, respectively.}
\label{fig:ksne-B-2d-regression}
\end{figure}

In the regression model (\autoref{fig:ksne-B-2d-regression}a) for the KS subpopulation (\autoref{fig:ksne-selection}a), \textbf{solar radiation} is significantly correlated with \textbf{growth rate} under the significant-level of $0.05$, whereas \textbf{humidity} is at the margin of statistical significance with a p-value $\leq 0.07$ (blue boxes in \autoref{fig:ksne-B-2d-regression}a).  
On the other hand, for the NE subpopulation (\autoref{fig:ksne-selection}b), only \textbf{solar radiation} is shown to be significantly correlated with the \textbf{growth rate} with a p-value $\leq 0.05$ (blue box in \autoref{fig:ksne-B-2d-regression}b). 
We obtained different regression models from these two selected subpopulations, meaning that the environmental variables have different influences on plants from the same genotype but grown in different locations.

\begin{figure}[!ht]
\centering
\includegraphics[width=0.55\columnwidth]{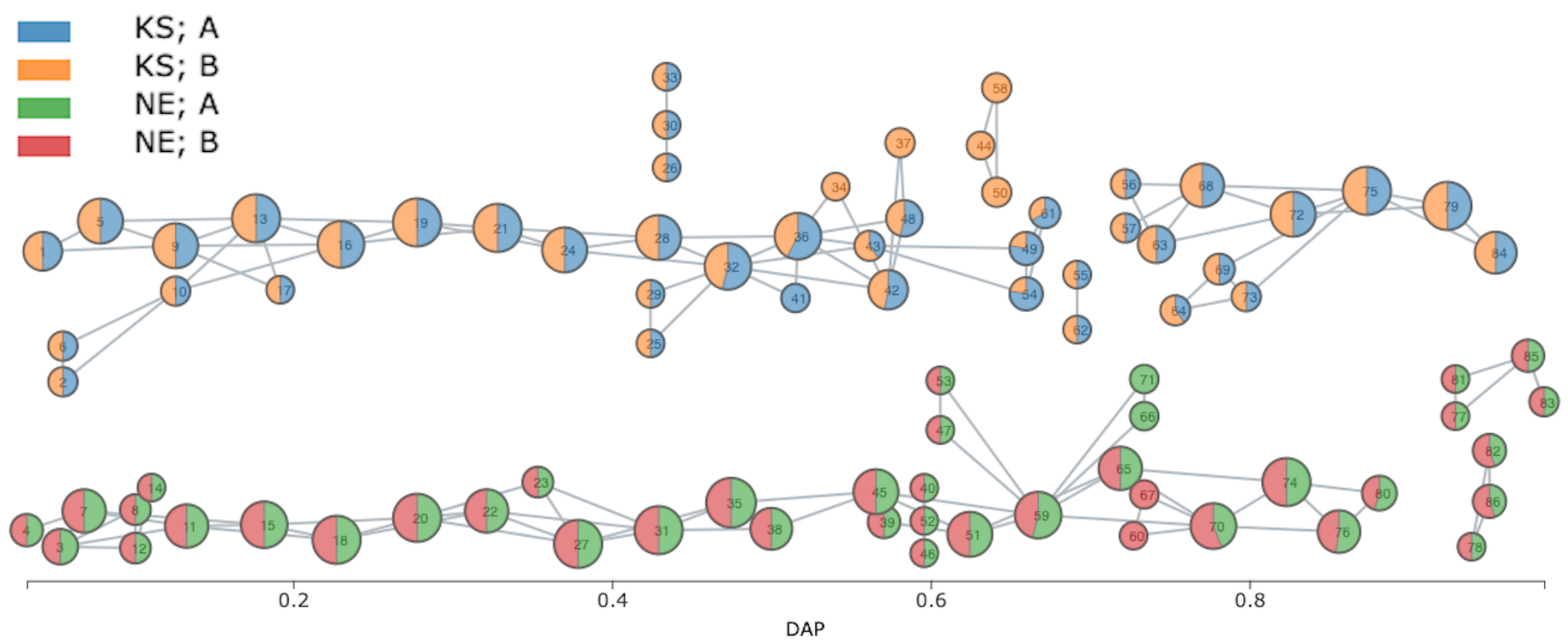}
\vspace{-4mm}
\caption{KS/NE dataset: the 1D mapper graph constructed with a 3D  (i.e, growth rate, solar radiation, and humidity) point cloud. Filter function: DAP. For DBSCAN, we chose $\epsilon = 0.15$,  $minPts = 2$. For mapper graph, we set $n=20$, $p=65\%$.}
\label{fig:ksne-high-dim}
\end{figure}

\para{Encoding environmental variables as additional dimensions.}
 Finally, we include the environmental variables, \textbf{solar radiation}  and \textbf{humidity}, together with \textbf{growth rate}, to form a 3D  point cloud for the mapper framework. 
We construct a 1D mapper graph using \textbf{DAP} as the filter function. The resulting mapper graph is shown in~\autoref{fig:ksne-high-dim}. 
The plants are now separated perfectly by location  starting from the beginning of planting. 
However, compared to the mapper graph of~\autoref{fig:ksne-time}, this mapper graph does not adequately distinguish the two genotypes (\textbf{A} vs \textbf{B}).  
 
\begin{figure}[!ht]
\centering
\includegraphics[width=0.35\columnwidth]{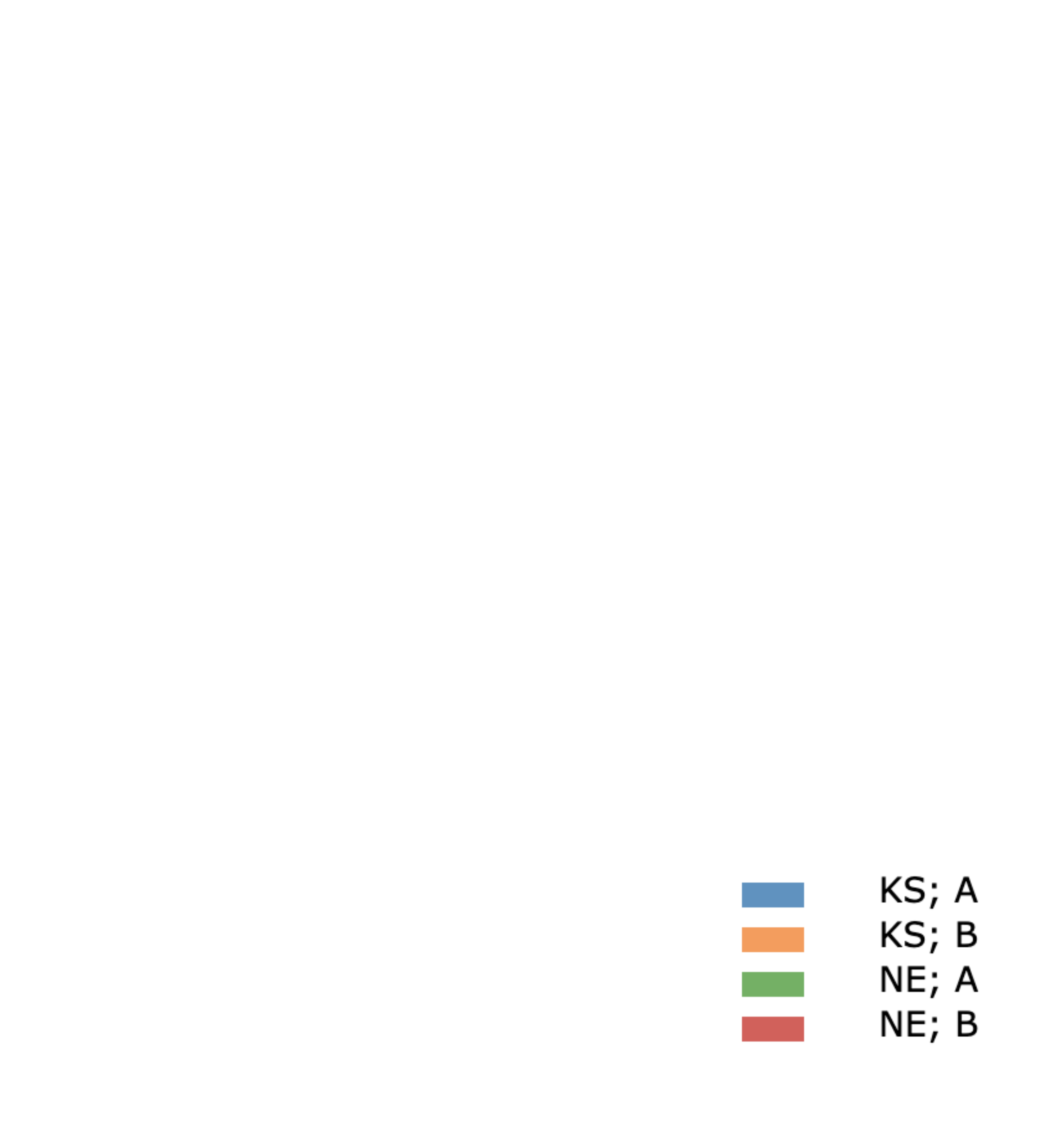}
\vspace{-4mm}
\caption{KS/NE dataset: t-SNE embedding of the 3D point cloud colored by the four (location, phenotype) categories.}
\label{fig:ksne-tsne}
\end{figure}

The above observation indeed aligns with our analysis of this 3D point cloud data using the built-in dimensionality reduction module. 
As shown in \autoref{fig:ksne-tsne}, we see that a t-SNE embedding of this point cloud clearly shows the separation by location. 

\begin{figure}[!ht]
\centering    
\includegraphics[width=0.8\columnwidth]{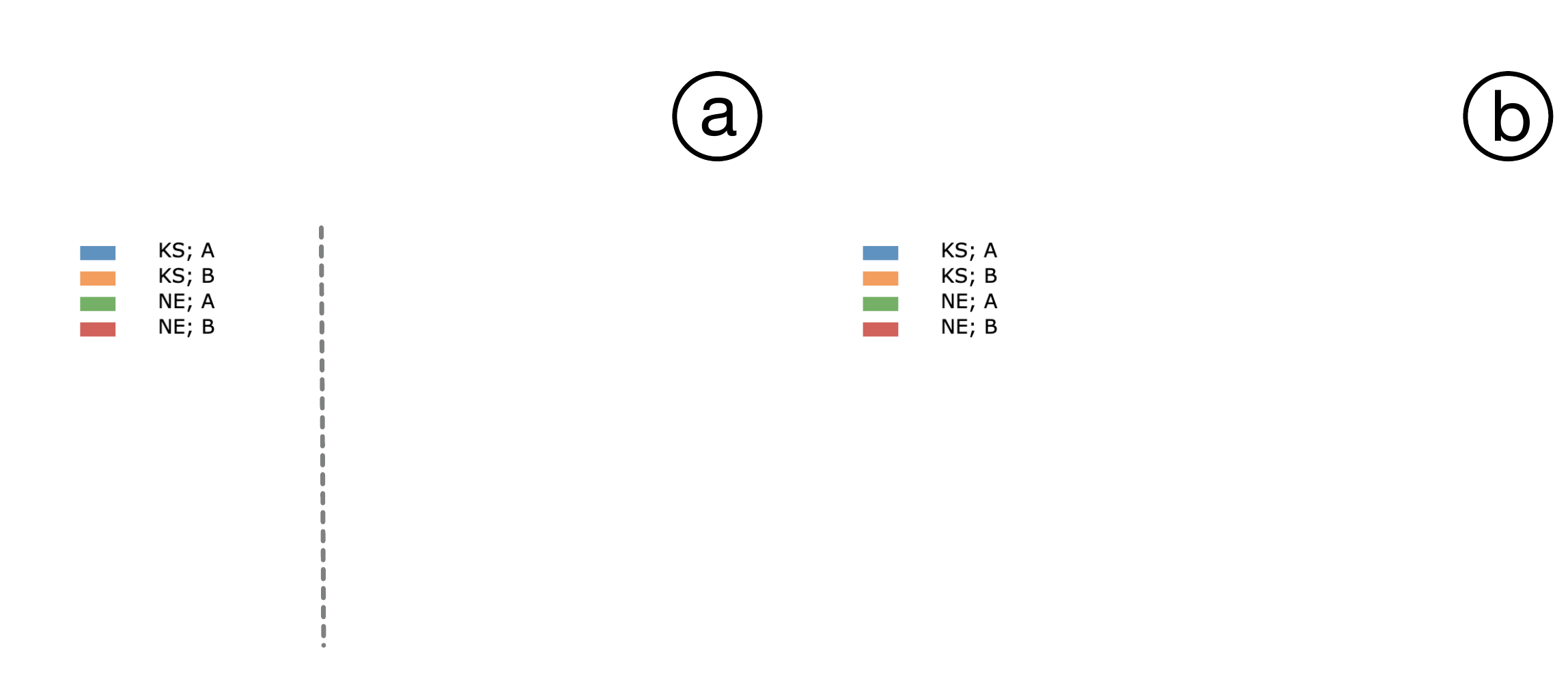}
\vspace{-4mm}
\caption{KS/NE dataset: (a) scatter plot of humidity vs. growth rate; (b) scatter plot of \textbf{DAP} vs. \textbf{growth rate}. Points are colored by the four (location, genotype) categories.}
\label{fig:ksne-scatter-plots}
\end{figure}

\para{Scatter plot analysis.}
To investigate such an observation further, we utilize the scatter plot modules provided by {\tool}. 
As shown in~\autoref{fig:ksne-scatter-plots}a, the scatter plot of \textbf{humidity} (x-axis) vs. \textbf{growth rate} (y-axis) shows a clear separation between KS subpopulations and NE subpopulations by location (see the grey dotted separating boundary). 
The same is true for the scatter plot of \textbf{solar radiation} vs. \textbf{growth rate} (not shown here). 
The scatter plot, again, confirms that the environmental variable \textbf{humidity} (or \textbf{solar radiation}) alone differentiate the plants by their planting locations.  
 
We push our exploratory analysis further by studying \textbf{DAP} vs \textbf{growth rate} using the scatter plot module. 
We observe a unimodal distribution for each of the four (location, phenotype) categories, as shown in \autoref{fig:ksne-scatter-plots}b, which is quite interesting.

\section{Use Case: Irrigation Dataset}
\label{sec:irrigation}

For the second maize dataset, referred to as the \textbf{Irrigation} dataset, data are collected from two field locations in NE with identical conditions except for the \textbf{irrigation} environmental variable: one location was irrigated but the other was not; this dataset has been explored previously by Kamruzzaman \etal~\cite{KamruzzamanKalyanaramanKrishnamoorthy2019}.  
The measured maize plants have more biodiversity since they come from 80 different genotypes.  
Similar to the KS/NE dataset, the columns consist of the genotype of each plant (among 80 genotypes), \textbf{DAP}, its growth information such as the \textbf{height difference} and the \textbf{growth rate difference}, and the weather information of each day, such as the \textbf{temperature}, \textbf{humidity}, etc. 
The dataset contains 6400 rows (data points). 

\subsection{Reproducing Known Results}

To reproduce the experimental results of Kamruzzaman \etal~\cite{KamruzzamanKalyanaramanKrishnamoorthy2019}, we use the \textbf{growth rate difference} to form our point cloud and \textbf{DAP} as the filter function to construct a 1D mapper graph. 

\begin{figure}[!ht]
\centering
\includegraphics[width=0.55\columnwidth]{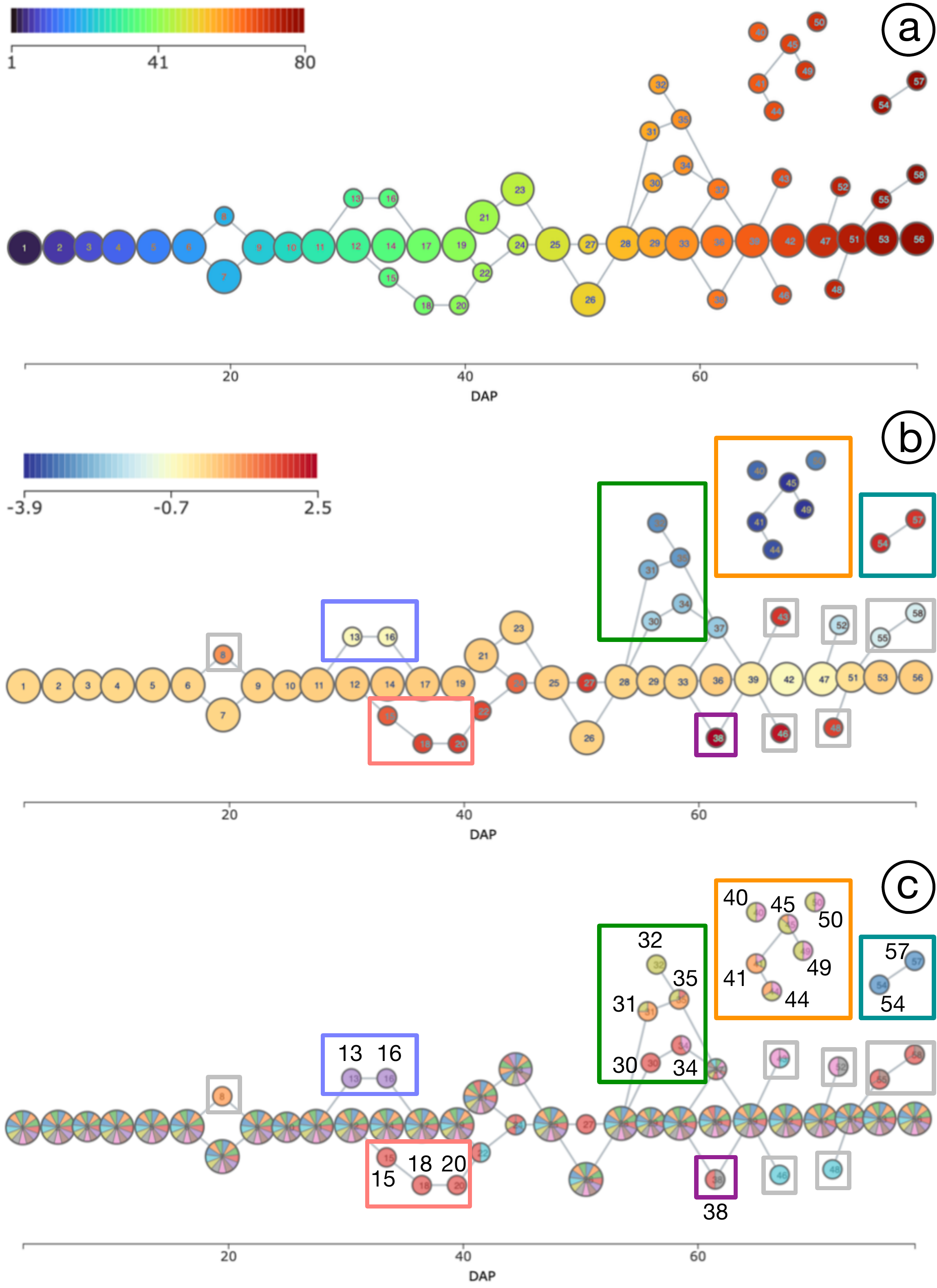}
\vspace{-4mm}
\caption{Mapper graphs of the Irrigation dataset aligned along the horizontal time axis quantified by DAP: (a) nodes are colored by DAP; (b) nodes are colored by growth rate difference; (c) nodes are visualized with pie charts according to $80$ genotypes. Pie charts involving more than 16 categories are visualized with a fixed glyph. For DBSCAN, we chose $\epsilon = 0.15$,  $minPts = 2$. For mapper graph, we set $n=28$, $p=27\%$.}
\label{fig:irrigation-time}
\end{figure}

We again work with an alternative mapper graph layout that aligns \textbf{DAP} along the x-axis, as shown in~\autoref{fig:irrigation-time}. 
We are interested in identifying genotypes that have varying \textbf{growth rate differences} between the two field locations; such genotypes are considered as \emph{anomalies} in the population. 
As shown in~\autoref{fig:irrigation-time}c, we observe that at nodes 13 (${\sim}$30 days) and 16 (${\sim}$34 days), the genotype \textit{PHW52 x LH123HT} bifurcates from other genotypes (blue box). 
At nodes 15 (${\sim}$34 days), 18, and 20 (${\sim}$40 days), the genotype \textit{PHB47 x PHR55} bifurcates from other genotypes (red box). 
Among nodes 30, 31, 32, 34, and 35 (${\sim}$ 55 - 59 days, green box),  four genotypes bifurcate from other genotypes, including \textit{LH198 x PHW30}, \textit{PHW52 x Q381}, \textit{PHB47 x PHG83}, and \textit{LH198 x LH51}.
At node 38 (${\sim}$ 62 days, purple box), the genotypes \textit{PHB47 x LH185} and \textit{PHP02 x PHB47} bifurcate from other genotypes.
Among nodes 40, 41, 44, 45, 49, and 50 (${\sim}$ 65 - 70 days, orange box), three genotypes bifurcate from other genotypes, including \textit{PHB47 x PHG83}, \textit{LH198 x LH51}, and \textit{PHB47 x LH38}.
Finally, at nodes 54 (${\sim}$75 days) and 57(${\sim}$79 days), the genotype \textit{ICI 441 x PHZ51} bifurcates from other genotypes (teal box). 
The same anomaly detection applies to the nodes enclosed by gray boxes as well. 
As shown in~\autoref{fig:irrigation-time}b, these genotypes/subpopulations  enclosed by colored boxes have distinct \textbf{growth rate differences}, and thus are considered as anomalies based on our topological analysis. 
 
This result demonstrates that using {\tool}, we are able to reproduce the  insights regarding when specific genotypes with different growth rates start to deviate from the main population of plants. 
Such subpopulations of anomalies can be further used to study the characteristics of these specific genotypes. 

\subsection{Conducting New In-depth Analysis}
With {\tool}, we can now perform new, in-depth analysis of selected subpopulations. 
Using the Irrigation dataset, we demonstrate the analysis and visualization capabilities of {\tool} in studying subpopulations. 
Specifically, we perform linear regression on the entire population of 80 genotypes as well as selected abnormalities as shown in~\autoref{fig:irrigation-time}c.

\begin{figure}[!ht]
\centering
\includegraphics[width=0.65\columnwidth]{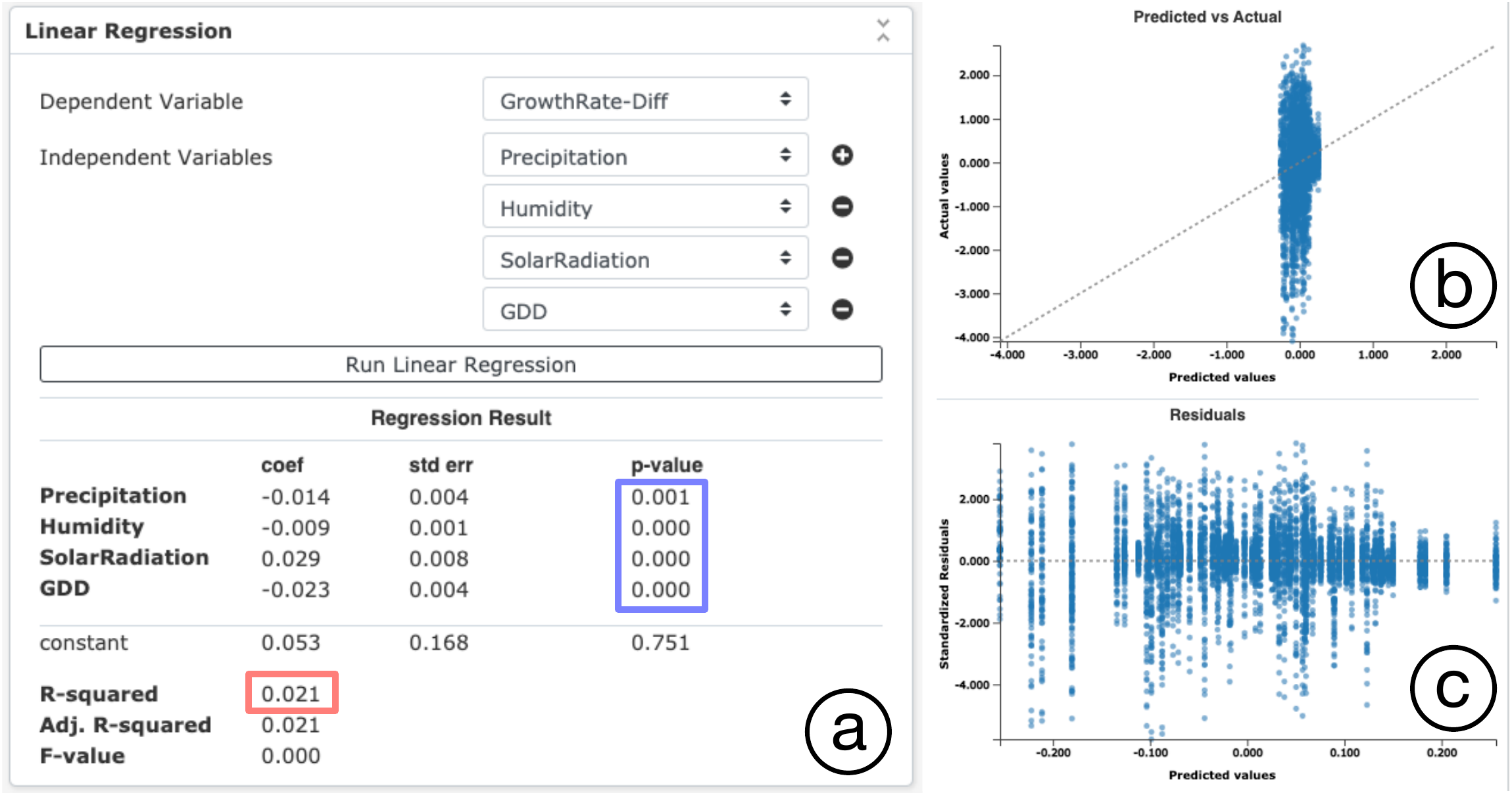}
\vspace{-2mm}
\caption{Irrigation dataset: (a) linear regression on the entire population; (b) predicted values vs. actual values plot; (c) residual plot of the model.}
\label{fig:irrigation-regression}
\end{figure}

\para{Linear regression on the entire population.}
We first perform a linear regression on the entire population of the Irrigation dataset. For this dataset, we treat \textbf{growing degree days} (GDD) as a temperature measurement when plants show a phenotypic response. As shown in~\autoref{fig:irrigation-regression}a, all the environmental variables are significantly related to the \textbf{growth rate difference} based on the p-values (blue box). 
However, the R-squared value ($0.021$, red box) is low, which informs us of the relative low predictive capability of the model based on these variables (see~\cite{LewisBeckSkalaban1990} for an interpretation of the R-squared). 
We extend the tool by adding the predicted values versus actual values plot and the residual plot under the regression result panel; see  ~\autoref{fig:irrigation-regression}b-c, which further confirm the low predictive capability of the model. 

\para{Linear regression on selected subpopulations.}
We are also interested in studying how environmental variables affect the   plants of different genotypes. 
We study a few \emph{abnormal} subpopulations with \textbf{growth rate differences} that are noticeably distinct from the rest of the population, such as those enclosed by colored boxed in~\autoref{fig:irrigation-time}b-c.   

\begin{figure}[!ht]
\centering
\includegraphics[width=0.55\columnwidth]{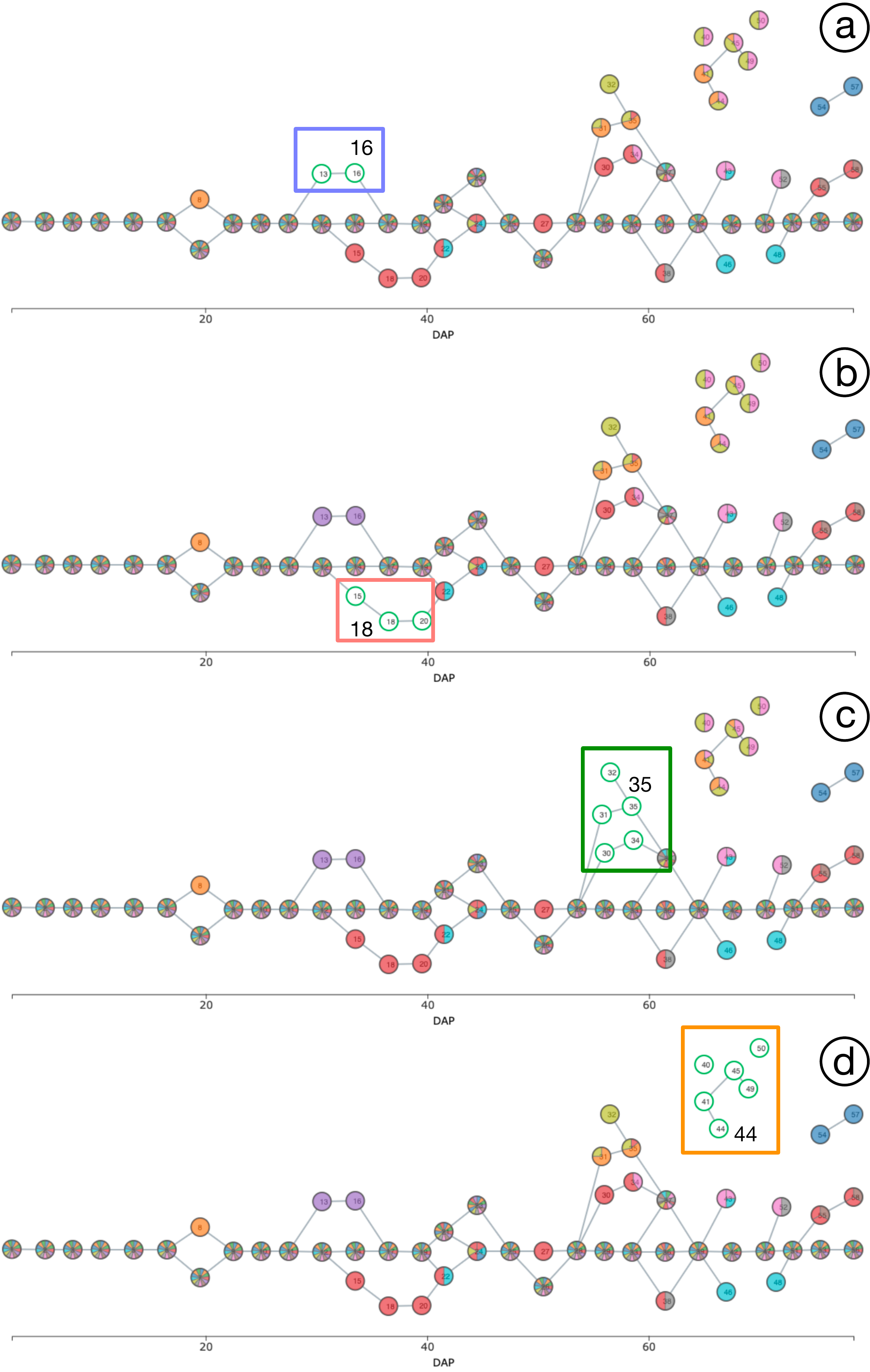}
\vspace{-4mm}
\caption{Irrigation dataset: green nodes in (a)-(d) correspond to the selected subpopulations for linear regression.}
\label{fig:irrigation-selection}
\end{figure}

\begin{figure}[!ht]
\centering
\includegraphics[width=0.65\columnwidth]{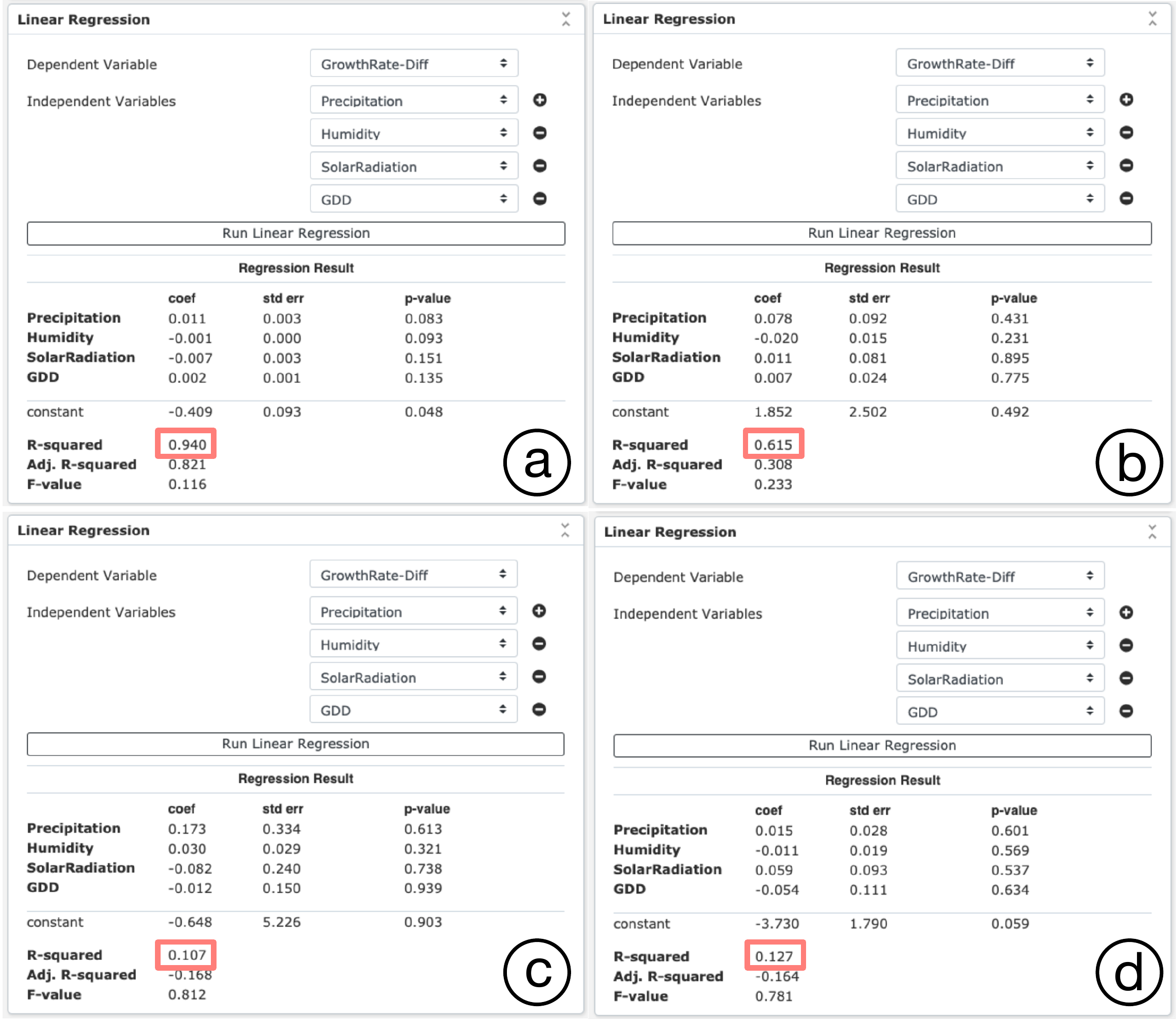}
\vspace{-2mm}
\caption{Irrigation dataset: linear regression for selected subpopulations.}
\label{fig:irrigation-local-regression}
\end{figure}

These selected subpopulations are shown in~\autoref{fig:irrigation-selection}. 
We now apply linear regression to these subpopulations. 
Recall from the previous section, the subpopulation (blue box in \autoref{fig:irrigation-selection}a) containing nodes 13 and 16 is related to genotype \emph{PHW52 x LH123HT}.
The subpopulation containing nodes 15, 18, and 20 (red box in \autoref{fig:irrigation-selection}b) is related to genotype \emph{PHB47 x PHR55}. 
Linear regression applied to these subpopulations shows relatively high R-squared values, as shown in~\autoref{fig:irrigation-local-regression}a and \autoref{fig:irrigation-local-regression}b (red boxes), respectively. 

However, for the subpopulation (green box in \autoref{fig:irrigation-selection}c) containing nodes 31, 32, 34, and 35 with four genotypes, and the subpopulation containing nodes 40, 41, 44, 45, 49, and 50 (orange box in \autoref{fig:irrigation-selection}d) with three genotypes, the R-squared values are both relatively low, as shown in~\autoref{fig:irrigation-local-regression}c-d (red boxes), respectively. 
 
The different regression results for these four subpopulations indicate that the environment variables tend to have different effects on plants from different genotypes with abnormal behaviors.

\begin{figure}[!ht]
\centering
\includegraphics[width=0.65\columnwidth]{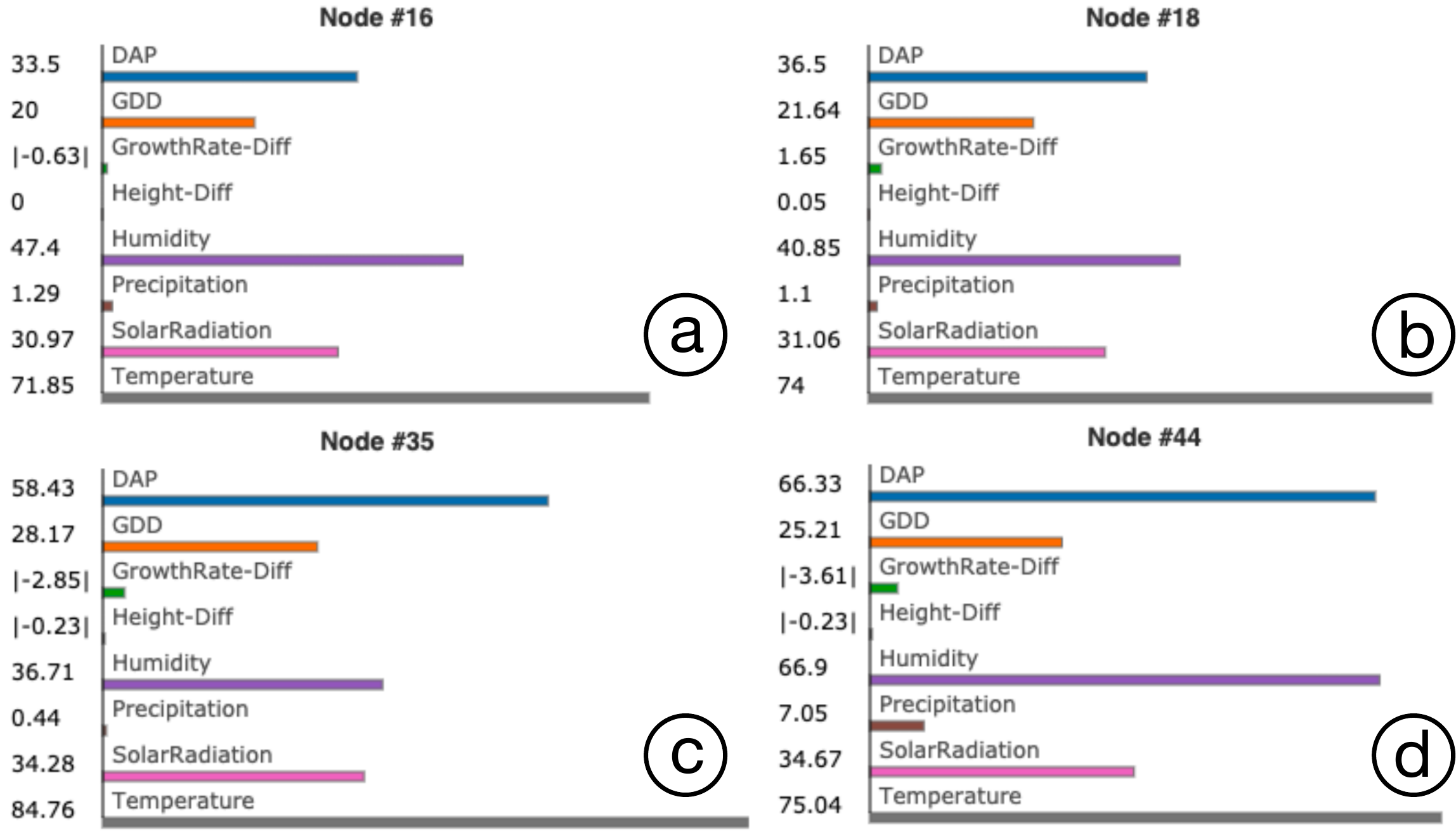}
\vspace{-2mm}
\caption{Irrigation dataset: the bar charts of selected subpopulations.}
\label{fig:irrigation-bar-charts}
\end{figure}

\para{Bar charts of selected nodes.}
For each selected node in a mapper graph, {\tool} provides a bar chart that treats each dimension (column) as a separate category, and represents the average values of each dimension with rectangular bars. 
We utilize these bar charts to further explore how selected nodes/genotypes differ from one another in terms of the environmental conditions. 
As shown in~\autoref{fig:irrigation-bar-charts}, we select nodes from each subpopulation of abnormalities (\autoref{fig:irrigation-selection}b-d) and observe the differences among the average environmental variables of plants contained in these nodes, such as the average \textbf{temperature}, \textbf{humidity}, \textbf{precipitation}, and \textbf{solar radiation}. 

\begin{figure}[!ht]
\centering
\includegraphics[width=0.5\columnwidth]{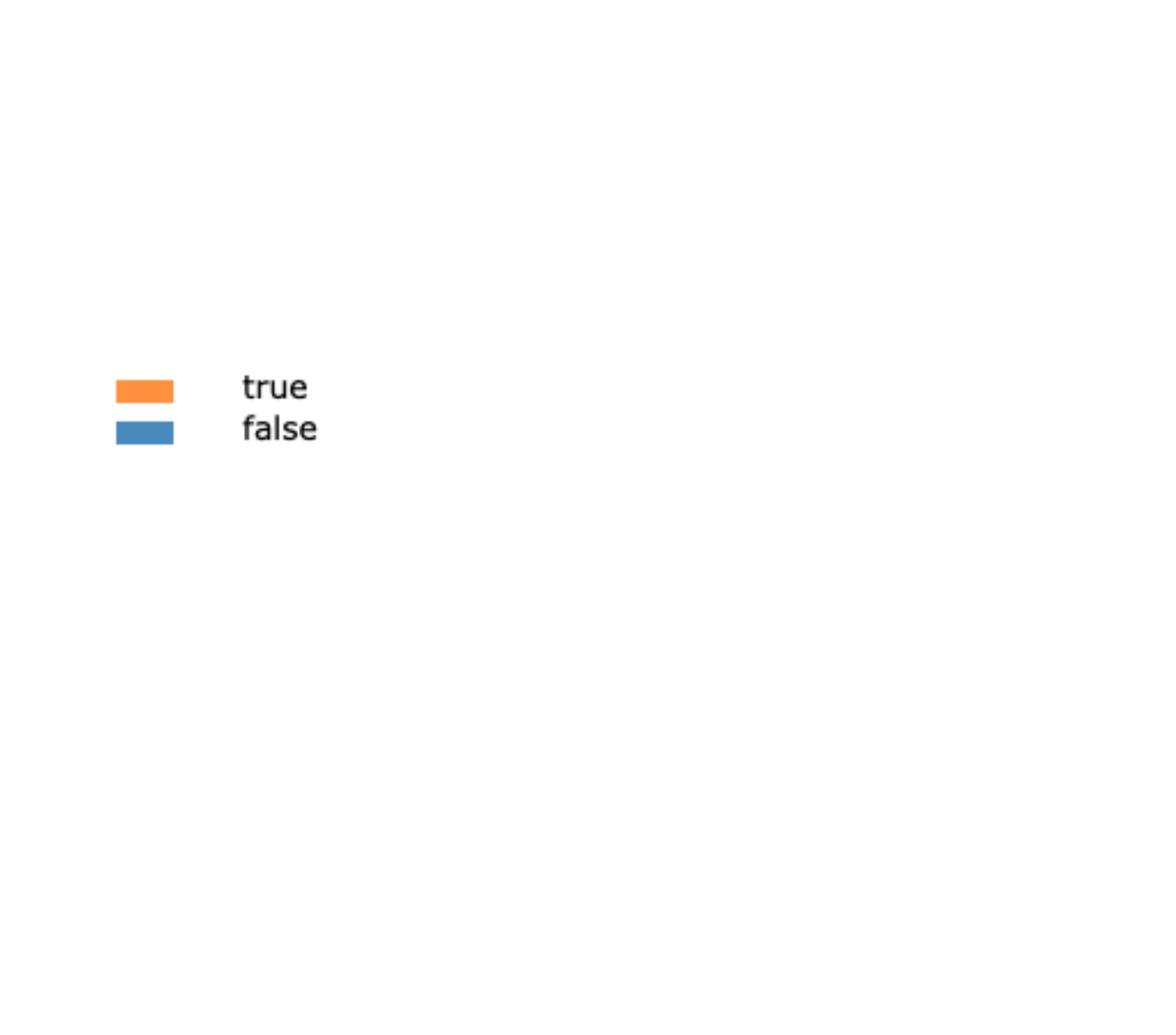}
\vspace{-2mm}
\caption{Irrigation dataset: the scatter plot of growth rate difference vs. DAP. Anomalies are colored orange while the rest of the population are colored blue.}
\label{fig:irrigation-scatter-plot}
\vspace{-4mm}
\end{figure}

\para{Scatter plot analysis of abnormal subpopulations.}
Finally, we apply scatter plot analysis of the abnormalities identified using   the mapper graph based analysis (as seen in~\autoref{fig:irrigation-time}c). 
As shown in \autoref{fig:irrigation-scatter-plot}, as time (\textbf{DAP}) increases, different genotypes have increased diversity in terms of their \textbf{growth rate differences}. 
At the same time, the anomalies identified by our topological approach are now shown to be on the boundaries of the scatter plot. 
The scatter plot indicates that the mapper graph helps to identify ``extremities'' as anomalies in this setting. 

\section{Conclusion}
\label{sec:conclusion}

We presented {\tool}, an interactive visualization tool for complex multidimensional phenomics data. 
Notably, the tool provides the user a way to interactively explore the data and perform in-depth analysis by visually exploring and comparing subpopulations. 
For example, {\tool} contains a number of built-in data analysis and ML modules useful for feature selection, linear regression, and detection of anomalous subpopulations. 
These capabilities enable users to answer important questions such as \emph{``Which subsets of my population phenotypically differ from one another?}'',  \emph{``Which subsets of environment variables best correlate with those phenotypic changes?}'', ``\emph{Which genotypes display more plasticity than the others?}'', and so on. 
The ability to answer such questions is central to hypothesis formulation, which is, in particular, very challenging for multidimensional phenomics data.
{\tool} provides interactive and ML capabilities towards this direction. 

Furthermore, our case studies have not substantively examined the effects of different environments on given genotypes at the same  developmental stage (i.e., $G \times E$ interactions). 
To do this, we would need to collect data for the same genotypes from multiple (ideally many) locations and/or years; this would make for an excellent follow-up study.


\section*{Acknowledgement}
This work was in parts funded by U.S. National Science Foundation (NSF) awards DBI-1661375 and DBI-1661348. 



\begin{thebibliography}{10}

\bibitem{alkhalifah2018maize}
N.~AlKhalifah, D.~A. Campbell, C.~M. Falcon, J.~M. Gardiner, N.~D. Miller,
  M.~C. Romay, R.~Walls, R.~Walton, C.-T. Yeh, M.~Bohn, et~al.
\newblock Maize genomes to fields: 2014 and 2015 field season genotype,
  phenotype, environment, and inbred ear image datasets.
\newblock {\em BMC research notes}, 11(1):1--5, 2018.

\bibitem{BiasottiGiorgiSpagnuolo2008}
S.~Biasotti, D.~Giorgi, M.~Spagnuolo, and B.~Falcidieno.
\newblock {Reeb} graphs for shape analysis and applications.
\newblock {\em Theoretical Computer Science}, 392:5--22, 2008.

\bibitem{bush2012genome}
W.~S. Bush and J.~H. Moore.
\newblock Genome-wide association studies.
\newblock {\em PLOS Computational Biology}, 8(12):e1002822, 2012.

\bibitem{cairns2012maize}
J.~E. Cairns, K.~Sonder, P.~H. Zaidi, N.~Verhulst, G.~Mahuku, R.~Babu, S.~K.
  Nair, B.~Das, B.~Govaerts, M.~T. Vinayan, et~al.
\newblock Maize production in a changing climate: impacts, adaptation, and
  mitigation strategies.
\newblock {\em Advances in agronomy}, 114:1--58, 2012.

\bibitem{EsterKriegelSander1996}
M.~Ester, H.-P. Kriegel, J.~Sander, and X.~Xu.
\newblock A density-based algorithm for discovering clusters in large spatial
  databases with noise.
\newblock In {\em Proceedings of the 2nd International Conference on Knowledge
  Discovery and Data Mining}, pages 226--231. AAAI Press, 1996.

\bibitem{houle2010phenomics}
D.~Houle, D.~R. Govindaraju, and S.~Omholt.
\newblock Phenomics: the next challenge.
\newblock {\em Nature reviews genetics}, 11(12):855--866, 2010.

\bibitem{KalyanaramanKamruzzamanKrishnamoorthy2019}
A.~Kalyanaraman, M.~Kamruzzaman, and B.~Krishnamoorthy.
\newblock Interesting paths in the mapper complex.
\newblock {\em Journal of Computational Geometry}, 10(1):500--531, 2019.

\bibitem{kamruzzaman2018detecting}
M.~Kamruzzaman, A.~Kalyanaraman, and B.~Krishnamoorthy.
\newblock Detecting divergent subpopulations in phenomics data using
  interesting flares.
\newblock In {\em Proceedings of the 2018 ACM International Conference on
  Bioinformatics, Computational Biology, and Health Informatics}, pages
  155--164, 2018.

\bibitem{KamruzzamanKalyanaramanKrishnamoorthy2019}
M.~Kamruzzaman, A.~Kalyanaraman, B.~Krishnamoorthy, S.~Hey, and P.~Schnable.
\newblock {Hyppo-X}: A scalable exploratory framework for analyzing complex
  phenomics data.
\newblock {\em IEEE/ACM Transactions on Computational Biology and
  Bioinformatics}, 2019.

\bibitem{Kobourov2013}
S.~G. Kobourov.
\newblock Force-directed drawing algorithms.
\newblock In {\em Handbook of Graph Drawing and Visualization}, pages 383--408.
  Chapman and Hall/CRC, 2013.

\bibitem{kusmec2018harnessing}
A.~Kusmec, N.~de~Leon, and P.~S. Schnable.
\newblock Harnessing phenotypic plasticity to improve maize yields.
\newblock {\em Frontiers in plant science}, 9:1377, 2018.

\bibitem{lawrence2019idea}
C.~J. Lawrence-Dill, P.~S. Schnable, and N.~M. Springer.
\newblock Idea factory: the maize genomes to fields initiative.
\newblock {\em Crop Science}, 59(4):1406--1410, 2019.

\bibitem{LewisBeckSkalaban1990}
M.~S. Lewis-Beck and A.~Skalaban.
\newblock The {R}-squared: Some straight talk.
\newblock {\em Political Analysis}, 2:153--171, 1990.

\bibitem{LumSinghLehman2013}
P.~Y. Lum, G.~Singh, A.~Lehman, T.~Ishkanov, M.~Vejdemo-Johansson,
  M.~Alagappan, J.~Carlsson, and G.~Carlsson.
\newblock Extracting insights from the shape of complex data using topology.
\newblock {\em Scientific reports}, 3:1236, 2013.

\bibitem{madhobi2019visual}
K.~F. Madhobi, M.~Kamruzzaman, A.~Kalyanaraman, E.~Lofgren, R.~Moehring, and
  B.~Krishnamoorthy.
\newblock A visual analytics framework for analysis of patient trajectories.
\newblock In {\em Proceedings of the 10th ACM International Conference on
  Bioinformatics, Computational Biology and Health Informatics}, pages 15--24,
  2019.

\bibitem{SinghMemoliCarlsson2007}
G.~Singh, F.~M\'emoli, and G.~Carlsson.
\newblock Topological methods for the analysis of high dimensional data sets
  and {3D} object recognition.
\newblock In {\em Eurographics Symposium on Point-Based Graphics}, pages
  91--100, 2007.

\bibitem{tardieu2017plant}
F.~Tardieu, L.~Cabrera-Bosquet, T.~Pridmore, and M.~Bennett.
\newblock Plant phenomics, from sensors to knowledge.
\newblock {\em Current Biology}, 27(15):R770--R783, 2017.

\bibitem{VeenSaulEargle2019}
H.~J. van Veen, N.~Saul, D.~Eargle, and S.~W. Mangham.
\newblock {Kepler Mapper}: A flexible python implementation of the {Mapper}
  algorithm.
\newblock {\em Journal of Open Source Software}, 4(42):1315, 2019.

\bibitem{wray2013pitfalls}
N.~R. Wray, J.~Yang, B.~J. Hayes, A.~L. Price, M.~E. Goddard, and P.~M.
  Visscher.
\newblock Pitfalls of predicting complex traits from snps.
\newblock {\em Nature Reviews Genetics}, 14(7):507--515, 2013.

\bibitem{xu2016envirotyping}
Y.~Xu.
\newblock Envirotyping for deciphering environmental impacts on crop plants.
\newblock {\em Theoretical and Applied Genetics}, 129(4):653--673, 2016.

\bibitem{zhao2019crop}
C.~Zhao, Y.~Zhang, J.~Du, X.~Guo, W.~Wen, S.~Gu, J.~Wang, and J.~Fan.
\newblock Crop phenomics: current status and perspectives.
\newblock {\em Frontiers in Plant Science}, 10:714, 2019.

\bibitem{ZhouChalapathiRathore2021}
Y.~Zhou, {Nithin Chalapathi}, {Archit Rathore}, Y.~Zhao, and B.~Wang.
\newblock {Mapper Interactive}: A scalable, extendable, and interactive toolbox
  for the visual exploration of high-dimensional data.
\newblock In {\em {IEEE} 14th Pacific Visualization Symposium}, pages 101--110,
  2021.

\end{thebibliography}


\end{document}